\documentclass[12pt]{article}
\usepackage{amsmath,epsfig,amsfonts,fullpage,natbib,booktabs, graphicx,enumerate}
\usepackage[compact]{titlesec}
\usepackage[normalem]{ulem}
\usepackage[colorlinks=true,pdfstartview=FitV,linkcolor=blue,citecolor=blue,urlcolor=blue]{hyperref}
\newcommand{\var}{\textrm{var}}

\newcommand{\cov}{\textrm{cov}}

\newcommand{\vv}[1]{\mbox{\boldmath $#1$}}

\newcommand{\beq}{\begin{sxeqnarray*}}\newcommand{\eeq}{\end{eqnarray*}}
\newcommand{\beqn}{\begin{eqnarray}}\newcommand{\eeqn}{\end{eqnarray}}
\newcommand{\bit}{\begin{itemize}}\newcommand{\eit}{\end{itemize}}
\newcommand{\bnum}{\begin{enumerate}}\newcommand{\enum}{\end{enumerate}}
\renewcommand{\baselinestretch}{1.5}
\newtheorem{thm}{Theorem}[section]

\newtheorem{defn}{Definition}[section]
\newtheorem{lemma}{Lemma}[section]
\newtheorem{prop}{Proposition}[section]

\begin{document}

\section*{Statistical Inference for Generalized Integer\newline Autoregressive Processes}

\textbf{Pashmeen Kaur}$^{1,3}$ and
\textbf{Peter F. Craigmile}$^{2,4}$

$^1$
Department of Statistics,
The Ohio State University,
Columbus, OH 43210, USA

$^2$
Department of Mathematics and Statistics,
Hunter College,
New York, NY 10065, USA

$^3$\verb_kaur.138@osu.edu_ \;
$^4$\verb_peter.craigmile@hunter.cuny.edu_ \;

\textit{Last updated \today}
\vspace{-.2cm}

\begin{abstract}

A popular and flexible time series model for counts is the generalized integer autoregressive process of order $p$, GINAR($p$). These Markov processes are defined using thinning operators evaluated on past values of the process along with a discretely-valued innovation process. This class includes the commonly used INAR($p$) process, defined with binomial thinning and Poisson innovations. GINAR processes can be used in a variety of settings, including modeling time series with low counts, and allow for more general mean-variance relationships, capturing both over- or under-dispersion.  

While there are many thinning operators and innovation processes given in the literature, less focus has been spent on comparing statistical inference and forecasting procedures over different choices of GINAR process.
We provide an extensive study of exact and approximate inference and forecasting methods that can be applied to a wide class of GINAR($p$) processes with general thinning and innovation parameters. We discuss the challenges of exact estimation when $p$ is larger. We summarize and extend asymptotic results for estimators of process parameters, and present simulations to compare small sample performance, highlighting how different methods compare. We illustrate this methodology by fitting GINAR processes to a disease surveillance series.

\textbf{Keywords}: 
Markov time series processes for counts;
Thinning operators; 
Parameter estimation; 
Conditional maximum likelihood;
Forecasting.
\end{abstract}

\section{Introduction}
\label{sec:introduction}

Discrete-valued time series appear in many settings such as
epidemiology, social science, and finance.  While traditional time
series models such as the autoregressive integrated moving average
(ARIMA) processes can approximately explain such data, they do not
preserve discreteness and accurately fail to capture different
mean-variance relationships.  Also, when counts are low (e.g., when
dealing with incidence of rare diseases and events), ARIMA processes
provide a poor approximation to the distribution of the data.
Further, processes for continuous-valued time series fail to produce
integer forecasts, leading to the creation of arbitrary forecasting methods for discrete-valued series.

One popular class of processes for non-negative discrete-valued series, the
integer autoregressive process of order one (INAR(1)),  was introduced independently by \citet{mckenzie1985} and \citet{al1987first}.  The INAR(1) process can be considered the integer-valued analogue of the AR(1) process because they have a matching autocorrelation structure. This process was later extended to the order \textit{p} case by \citet{alzaid1990integer} and \citet{duli1991}, who defined the process using different dependence structures for the counting series. INAR($p$) processes are defined using a binomial thinning operation and a discretely-valued usually Poisson distributed innovation sequence, which guarantees that the series is discretely-valued.

While it is possible to generalize INAR($p$) processes to capture other marginal distributions (see e.g., \citet{mckenzie1985}, \citet{al1992first}, and \citet{ristic2012estimation} for extensions
to negative binomial marginal distributions), we can generalize further. We focus our attention on generalized
integer autoregressive models (GINAR($p$)), introduced by
\citet{latour1998existence}.  This class of models extends the
INAR($p$) process by allowing for different thinning operators.  The
inclusion of these thinning operators allows for more general time
series structure: we can further vary the choice of marginal
distributions, and can vary over- and under-dispersion structures
(e.g., \citet{jung2005estimation}, \citet{weiss2013integer}, \citet{bourguignon2015first}, \citet{huang2021new}.
(For a review of GINAR processes and other models for discrete time
series, also see \citet{weiss2008thinning}.   We also provide a literature review in the supplementary material.)

In this article we evaluate different methods of statistical inference for GINAR processes.  The most common parameter estimation methods are
conditional maximum likelihood (CML), Yule-Walker (YW) and conditional
least squares (CLS).  However, in many articles in the literature, focus is
given to the integer processes of order one, as processes with higher orders
can be more computationally intensive to estimate.
We focus on statistical inference for general orders $p$, not just
one, and provide asymptotic theory for estimating GINAR process
parameters with different estimation methods.  We will also compare how
well these methods perform using Monte Carlo simulations.  Further, we
investigate the construction of confidence regions for parameters,
highlight methods for model selection, and review and evaluate
methods for forecasting that preserve the discreteness of the
distribution.

The rest of this paper is organized as follows. Section \ref{sec:GINAR(p) Models} provides an introduction to thinning operators and provides the definition of the GINAR($p$) process. Section \ref{sec:est.methods} provides current estimation methodologies and extends them to the class of models defined in Section \ref{sec:GINAR(p) Models}. Section \ref{sec:Inference} discusses inference methodologies, and Section \ref{sec:sim.studies} provides a discussion of the simulation results comparing estimation methods for varying GINAR($p$) processes and model parameters. In Section \ref{sec:Application}, we apply these class of processes to a series of meningococcal disease cases in Germany. Lastly, Section \ref{sec:Conclusion and Discussion} presents a discussion of future work and summarizes the findings of this review. Other examples of GINAR processes, proofs of results, and supplementary simulations can be found in the supplementary material. R code that accompanies this article can be downloaded from \url{https://github.com/petercraigmile/GINAR/}.

\section{Generalized Integer Autoregressive (GINAR) processes}
\label{sec:GINAR(p) Models}

We begin by defining the generalized thinning operator which is an immediate extension of the binomial thinning operator~\citep{steutel1979discrete}. A thinning operator preserves the discreteness of the distribution.  

\begin{defn}
\label{gen.thin.def}
    Let $X$ be a non-negative integer-valued random variable (RV) with $\alpha \in [0,1]$. Then the generalized thinning operator, $\odot$, is defined by
\begin{align}
\label{eq:genthin}
    \alpha \odot X = \sum_{k=1}^{X} Y_k,
\end{align}
where $\{Y_k: k \in \mathbb{N}\}$ are a set of independent and
identically distributed (IID) RVs from some count distribution with
mean $\alpha$ and variance $\beta$~\citep{latour1998existence}. In
\eqref{eq:genthin} when $X = 0$, the sum is assumed to be 0.
\end{defn} 
We next define the two most popular thinning operators, binomial thinning and negative binomial thinning.  We provide other thinning operators in the supplemental material.

\textbf{Binomial thinning}: Suppose $Z = \{Y_k: k \in \mathbb{N}\}$ is a set of IID Bernoulli RVs with success probability $\alpha$, Bern($\alpha$). Then $\alpha \odot X$ given $X=x$ follows a binomial distribution with size $x$ and success probability $\alpha$, Binomial($x$, $\alpha$) with probability mass function (pmf) evaluated at $z$ of
\begin{align*}
    \binom{x}{z}  \alpha^z(1-\alpha)^{(x-z)}I\{0 \leq z \leq x\}.
\end{align*}

\textbf{Negative binomial thinning}: Let
$\{Y_k: k \in \mathbb{N}\}$ be a set of IID Geometric($1/(1+\alpha)$) RVs, with pmf given by $P(Y_k=y)= \alpha^y/(1+\alpha)^{y+1}I\{y \geq 0\} $. Then $\alpha \odot X$ given $X=x$ has a negative binomial distribution with size $x$ and success probability $1/(1+\alpha)$ when $x > 0$, and is 0 with probability 1 when $x=0$~\citep{ristic2009new}.

We can now define the class of GINAR($p$) process. 
\begin{defn}
\label{defn:ginarp}
The GINAR($p$) process~\citep{latour1998existence} $\{X_t : t \in \mathbb{Z}\}$ is the non-negative integer-valued stationary and ergodic process defined by
\begin{align}
\label{eq:ginarp}
    X_t = \sum_{j=1}^p \alpha_j \odot X_{t-j} + \epsilon_t, \quad t \in \mathbb{Z}.
\end{align}
In the above definition, $\alpha_j \in [0,1)$ for each $j=1,\ldots,p$
with $\sum_{j=1}^p \alpha_j < 1$, and the innovation process
$\{\epsilon_t: t \in \mathbb{Z}\}$ is a set of IID non-negative integer-valued RVs
with $\mu_\epsilon > 0 $ and variance $\sigma^2_\epsilon > 0$. Also, the $\{Y_k\}$ associated with each thinning operation (see (\ref{eq:genthin})) are mutually independent and independent of $\{\epsilon_t\}$, and $\epsilon_t$ is independent of $X_{t-j}$ for all $t \in \mathbb{Z}$ and $j \geq 1$.
\end{defn}

When we use binomial thinning in (\ref{eq:ginarp}), we obtain the INAR($p$) process of \citet{duli1991} with the INAR(1) process of \citet{al1987first} and \citet{mckenzie1985} when $p=1$. The less commonly used INAR($p$) process of \citet{al1992first} cannot be written in this form, and we do not consider this process further.

\subsection{Statistical properties}
\label{stat.properties}

In this section we present useful statistical properties of both the generalized thinning operator and the GINAR(\textit{p}) process.

\begin{lemma}\label{lemma:lemma1}
  The thinning operator defined by Definition~\ref{gen.thin.def} has the following properties.
  \bnum[(a)]
\item $0 \odot X = 0$;
  \item $1 \odot X = X$;
\item $E(\alpha \odot X) = \alpha E(X)$;
\item  $E(\alpha \odot X)^2 = \alpha^2 E(X^2) + \beta E(X)$.
  \enum
\end{lemma}
It can be shown that for binomial and negative binomial thinning that $\beta = \alpha(1-\alpha)$ and $\beta = \alpha(1+\alpha)$, respectively. 
\begin{lemma}
  Let $\{X_t:t \in \mathbb{Z}\}$ be a GINAR($p$) process of Definition \ref{defn:ginarp}. Then for each $t \in \mathbb{Z}$, the conditional mean is
\begin{align*}
  \mu_{X_t|X_{t-1},...,X_{t-p}} = E(X_t|X_{t-1},...,X_{t-p}) &= \sum_{j=1}^p \alpha_j X_{t-j} + \mu_\epsilon,
\end{align*}
and the conditional variance is 
\begin{align*}
    \sigma^2_{X_t|X_{t-1},...,X_{t-p}} = \var(X_t|X_{t-1},...,X_{t-p}) &= \sum_{j=1}^p \beta_j X_{t-j} + \sigma^2_\epsilon.
\end{align*}
Further, the marginal mean and variance are respectively
\begin{eqnarray*}
  \mu_X \;=\; \frac{\mu_\epsilon }{ 1-\sum_{j=1}^p \alpha_j }
\quad{\mbox{and}}\quad
    \var(X_t) \;=\; \sum_{j=1}^p \beta_j \mu_X + \frac{\sigma^2_\epsilon} {1-\sum_{j=1}^p \alpha_j   \rho_X(j)},
\end{eqnarray*}
where $\{ \rho_X(k) : k \in \mathbb{Z} \}$ is the autocorrelation sequence for $\{X_t\}$.
The autocovariance sequence $\{ \gamma_X(k) : k \in \mathbb{Z} \}$ satisfies the following for all lags $k \neq 0$:
\begin{align*}
  \gamma_X(k) &= \alpha_1 \gamma_X(|k|-1) + \alpha_2 \gamma_X(|k|-2)+...+\alpha_p \gamma_X(|k|-p).
\end{align*}
Similarly, the autocorrelation sequence satisfies for all lags $k \neq 0$:
\begin{align*}
  \rho_X(k) &= \alpha_1 \rho_X(|k|-1) + \alpha_2 \rho_X(|k|-2)+...+\alpha_p \rho_X(|k|-p).
\end{align*}
The spectral density function, $f_X(\cdot)$, of the process is 
\begin{align*}
    f_X(\nu) = \frac{\sigma^2_\epsilon + \mu_X \sum_{j=1}^p \beta_j} {2\pi |\alpha(e^{-i\nu})|^2}, \quad \nu \in [-\pi, \pi],
\end{align*}
where $\alpha(B) = 1 - \sum_{j=1}^p \alpha_jB^j$ is a transfer function and $\beta$ is the backshift operator.
\label{lemma:lemma2}
\end{lemma}

Following Definition \ref{defn:ginarp}, the GINAR($p$) process is a Markov chain. The transition probabilities are given in the next theorem. (The proof follows directly by considering the convolutions that define the process.)

\begin{thm}
\label{thm:trans.prob}
\citep[][Theorem 1]{hadri2009maximum}
    For $\{X_t:t \in \mathbb{Z}\}$ a  GINAR($p$) process as defined in Definition \ref{defn:ginarp}, the transition probabilities are given by
\begin{eqnarray*}
&&\hspace{-1cm}
        P(X_t=x|X_{t-1}=x_{t-1},...,X_{t-p}=x_{t-p}) 
\\&=& 
    \sum_{i_1 = 0}^x P(\alpha_1 \odot x_{t-1}=i_1|X_{t-1}=x_{t-1})  \times
\\&&\hspace{0.5cm}
   \sum_{i_2=0}^{x-i_1} P(\alpha_2 \odot x_{t-2}=i_2|X_{t-2}=x_{t-2}) \times \ldots
    \times 
\\&&\hspace{1cm}
    \sum_{i_p=0}^{x-(i_1+i_2+...+i_{p-1})} P(\alpha_p \odot x_{t-p}=i_p|X_{t-p}=x_{t-p})  \times 
\\&&\hspace{3cm}
     P(\epsilon_t = x - (i_1 + i_2+...+i_p)).
\end{eqnarray*}
\end{thm}

For larger values of $p$ the calculation of these convolutions become computationally burdensome (on average $O(\mu_X^p)$, where $\mu_X$ is the mean of the process). Instead, noticing that convolutions can be rewritten as products in the Fourier domain, \cite{joe2019likelihood} proposes calculating the transition probabilities by integrating the characteristic function (chf) (e.g., \citet{davies1973numerical}). This leads to a more computationally efficient algorithm for calculating transition probabilities.

\begin{prop}
\label{prop:davies.tp}
For $\{X_t:t \in \mathbb{Z}\}$ a GINAR($p$) process as defined in Definition \ref{defn:ginarp}, suppose that $\phi_{X_t|X_{t-1},\ldots,X_{t-p}}(u)$ is the chf for the transition probability defined in Theorem \ref{thm:trans.prob}. Then the cumulative distribution function is
\begin{align}
    a(x) = P(X_t<x|X_{t-1},\ldots,X_{t-p}) = \frac{1}{2} - \frac{1}{2\pi} \int_{-\pi}^{\pi} Re \left(\frac{\phi_{X_t|X_{t-1},\ldots,X_{t-p}}(u) e^{-iux}}{1- e^{-iu}} \right) \,du,
\end{align}
where
\begin{align*}
\phi_{X_t|X_{t-1},\ldots,X_{t-p}}(u) = \phi_{\epsilon_t}(u) \prod_{j=1}^p \left[\phi_{Y^{(j)}}(u)\right]^{X_{t-j}}.
\end{align*}

Then the transition probabilities can be calculated 
\begin{align*}
b(x) = P(X_t=x|X_{t-1},\dots,X_{t-p}) =
   \left\{
   \begin{array}{ll}
   a(1),         & x=0;\\
   a(x+1) - a(x), & x=1,2,\ldots,
   \end{array}
   \right.
\end{align*}
where
\begin{align*}
    b(x) &= \frac{1}{\pi} \int_{0}^{\pi} \mbox{Re}\!\left( \phi_{X_t|X_{t-1},\ldots,X_{t-p}}(u) \; e^{-iux_t}  \right) \,du, \quad x=1,2, \ldots.
\end{align*}
\end{prop}

\section{Estimation methods}
\label{sec:est.methods}
Let $\boldsymbol{X}=(X_1,\ldots, X_n)^T$ be observations from a GINAR($p$) process where the thinning operator and innovation process is known. We are interested in estimation of the parameter vector $\boldsymbol{\theta}=(\alpha_1, \alpha_2,...,\alpha_p, \mu_{\epsilon}, \sigma^2_{\epsilon})^T$, where $\sum_{j=1}^p \alpha_j < 1$ and the constraints on $\mu_{\epsilon}$ and $\sigma^2_{\epsilon}$ depend on the form of the innovation distribution. Let $\boldsymbol{\Theta}$ be the resulting parameter space for $\boldsymbol{\theta}$, which we assume to be compact. We further assume that our GINAR($p$) process is identifiable; we can tell apart different values of the parameter vector on the basis of the transition probabilities.

In this section we discuss the following methods: conditional maximum likelihood (CML), Yule-Walker (Y-W), pseudo maximum likelihood, conditional least squares (CLS), Whittle likelihood, and saddlepoint methods.

\subsection{Conditional maximum likelihood}
\label{est.methods.cml}

Conditioning on the first $p$ observations, CML calculates the conditional log likelihood using
\begin{align}
    \ell(\boldsymbol{\theta}) 
    &= \sum_{t=p+1}^n \log P(X_t = x_t|X_{t-1},..., X_{t-p}) \;=\; \sum_{t=p+1}^n \log b(x_t),
\end{align}
where $b(\cdot)$ is defined in Proposition \ref{prop:davies.tp}.
The CML-based parameter estimates of $\vv{\theta}$ can then be computed as
\begin{align}
\label{eq:argmax}
    \widehat{\boldsymbol{\theta}}_{CML} = \underset{\boldsymbol{\theta} \in \Theta}{\arg \max} \ \ell(\boldsymbol{\theta}).
\end{align}

Numerical optimization techniques are typically used to maximize \eqref{eq:argmax}.
 As discussed in Section \ref{stat.properties}, the form of the transition probabilities will depend on the thinning operation and marginal distribution of innovation. We know the computational complexity of CML increases with $p$, and a brute force method of coding the transition probabilities is not efficient. Thus we use Proposition \ref{prop:davies.tp} to efficiently evaluate transition probabilities using numerical quadrature techniques.

The following theorem shows that the CML estimator is asymptotically normal.

\begin{thm}
\label{thm:cml.asymptotics}
 Let $\{X_t:t \in \mathbb{Z}\}$ be a GINAR($p$) process. Then the conditional maximum likelihood estimator, $\widehat{\boldsymbol{\theta}}_{CML}$, has the following asymptotic distribution
\begin{equation}
    \sqrt{n}(\widehat{\boldsymbol{\theta}}_{CML}-\boldsymbol{\theta}) \xrightarrow{\text{d}} N(\boldsymbol{0}, \boldsymbol{\Sigma}_{CML}),
\end{equation}
 as $n \rightarrow \infty$, where 
 \begin{align*}
 \boldsymbol{\Sigma}_{CML} &= \mathbf{{J(\boldsymbol{\theta})}^{-1}K(\boldsymbol{\theta}){J(\boldsymbol{\theta})}^{-1}}, \\
     \mathbf{K(\boldsymbol{\theta})} &= \underset{n \rightarrow \infty}{lim} \ \frac{1}{n} E\left( \frac{\partial \ell(\boldsymbol{\theta})}{ \partial \boldsymbol{\theta}} \left({\frac{\partial \ell(\boldsymbol{\theta})}{ \partial \boldsymbol{\theta}}}\right)^T   \right),
\end{align*}
and
\begin{align*}
     \mathbf{J(\boldsymbol{\theta})} &= \underset{n \rightarrow \infty}{lim} \ \frac{1}{n} E\left(\frac{\partial^2 \ell(\boldsymbol{\theta})}{\partial \boldsymbol{\theta} \partial \boldsymbol{\theta}^T} \right).
 \end{align*}
\end{thm}

\citet[][Theorem 2 and Theoreom 3]{hadri2009maximum} provide expressions for first and second derivatives of the log likelihood function for a GINAR($p$) process and show calculations for the expectations of derivatives for the GINAR(2) process with binomial thinning and Poisson innovations. Alternatively, we can calculate these quantities using the chf, as follows.

\begin{prop}
\label{prop:cml.derivatives}
\begin{sloppypar}
In Proposition \ref{prop:davies.tp} we defined the transition probability $b(x) = P(X_t|X_{t-1},\ldots,X_{t-p})$ for $x=0,1,\ldots$.
Fixing $x$, let $b^{(j)}(x)$ denote the partial derivative of $b(x)$ with respect to $\theta_j$ and $b^{(j,k)}(x)$ denote the second partial derivative of $b(x)$  with respect to $\theta_j$ and $\theta_k$.  Then
\end{sloppypar}\vspace{-1cm}
\begin{eqnarray*}
    \frac{\partial}{\partial \theta_j} \ell(\boldsymbol{\theta}) 
\;=\;
    \sum_{t=p+1}^n \frac{ b^{(j)}(x_t) }{ b(x_t) },
\end{eqnarray*}
with
\begin{eqnarray*}
    \frac{\partial^2}{\partial \theta_j \partial \theta_k} \ell(\boldsymbol{\theta}) 
\;=\;
    \sum_{t=p+1}^n \frac{ b(x_t) \: b^{(j,k)}(x_t) - b^{(j)}(x_t) \: b^{(k)}(x_t) }{ b^2(x_t) }.
\end{eqnarray*}
\label{thm:Proposition 1}
\end{prop}

\subsection{Yule-Walker}

Yule-Walker (Y-W) estimation~\citep[e.g.,][Section 5.1.1]{Brock} is a method of moments approach. From Proposition \ref{stat.properties}, we see that the GINAR($p$) and AR($p$) processes have the same autocorrelation structure. Consequently the Y-W equations hold for the GINAR($p$) process:
\begin{align}
\label{eq:yw}
    \boldsymbol{\Gamma}\boldsymbol{\alpha} = \boldsymbol{\rho},
\end{align}
where $\boldsymbol{\Gamma} = [\gamma_X(|i-j|)]_{p\times p}, \ \
    \boldsymbol{\alpha} = [\alpha_1,...,\alpha_p]^T, \ \ \mbox{and }
    \boldsymbol{\gamma} = [\gamma_X(1),...,\gamma_X(p)]^T$.
Replacing the quantities in (\ref{eq:yw}) with the corresponding sample estimates, i.e. $\widehat{\boldsymbol{\Gamma}}\widehat{\boldsymbol{\alpha}}_{YW} = \widehat{\boldsymbol{\boldsymbol{\rho}}}$, provides the Y-W estimate for $\boldsymbol{\alpha}$. Here the sample autocovariance, $\widehat{{\gamma}}_X(k)$, is \begin{align*}
    \widehat{{\gamma}}_X(k) = \frac{1}{n} \sum_{t=1}^{n- \lvert k \rvert} (X_t - \overline{X})(X_{t+ \lvert k \rvert}-\overline{X}), \quad k \in \mathbb{Z}. 
\end{align*}
The Y-W estimates of $\mu_\epsilon$ and $\sigma^2_\epsilon$ are~\citep{silva2006asymptotic} 
\begin{align*}
\widehat{\mu}_{\epsilon} &= \left(1-\sum_{j=1}^p \widehat{\alpha}_j \right)\overline{X}
\end{align*}
and 
\begin{align*}
\sigma^2_\epsilon &= \widehat{V}_p - \overline{X}\sum_{j=1}^p \widehat{\beta}_j,
\end{align*}
where $\widehat{V}_p = \widehat{\gamma}_{\scriptscriptstyle X}(0) - \sum_{j=1}^p \widehat{\alpha}_j\widehat{\gamma}_{\scriptscriptstyle X}(j)$. 
The asymptotic distribution of $\widehat{\boldsymbol{\alpha}}_{YW}$ is as follows.
 
\begin{thm}
\label{yw.asymptotics}
\citep[][Theoreom 2]{silva2006asymptotic}
 Let $\{X_t:t \in \mathbb{Z}\}$ be a GINAR($p$) process. Assume that for $\{Y_k: k\in \mathbb{Z}\}$ defined in Definition \ref{gen.thin.def}, and $E(Y_k^3)$ that $E(Y_k^4)$ are finite for all $k$. Let $\widehat{\boldsymbol{\alpha}}_{YW}$ be the Y-W estimator of $\boldsymbol{\alpha}$. Then, 
\begin{align*}
    \sqrt{n}\left(\widehat{\boldsymbol{\alpha}}_{YW}-\boldsymbol{\alpha}\right) \xrightarrow{\text{d}} N\left(\boldsymbol{0}, \boldsymbol{\Sigma}_{YW}\right)
\end{align*}
 as $n \rightarrow \infty$, where $\boldsymbol{\Sigma}_{YW} = \mathbf{D}^T \mathbf{QD}$. In the definition of $\boldsymbol{\Sigma}_{YW}$, 
 \begin{align*}
 \boldsymbol{D}^T = -[\gamma_X(1)\boldsymbol{I}_p \dots  \gamma_X(p)\boldsymbol{I}_p](\boldsymbol{\Gamma}^T \otimes \boldsymbol{\Gamma}^{-1})[\boldsymbol{I}_{p^2} \ \boldsymbol{0}_{p^2 \times p}] + [\boldsymbol{0}_{p^2 \times p} \ \boldsymbol{\Gamma}^{-1}]
 \end{align*}
and $\boldsymbol{Q}$ is the $p(p+1) \times p(p+1)$ covariance matrix defined by
\begin{align*}
    \boldsymbol{Q} &= \cov(\widehat{V}_R(j),\widehat{V}_R(k)), \ \mbox{where } \\
    \widehat{V}_R(j) &= \begin{cases}
        \widehat{\gamma}_X(|(j-1) \ mod \ p-[(j-1)/p]|) & \text{if } j \leq p^2; \\
        \widehat{\gamma}_X(|j\ mod\ p|) & \text{if } j>p^2.
    \end{cases}
\end{align*}
Note that $\otimes$ is the Kronecker product \citep{graham1981kronecker}.
\end{thm}

The asymptotic distribution for  $\widehat{\sigma}^2_{\epsilon}$ and $\widehat{\mu}_{\epsilon}$ follows from 
\citet[][Theorem 1, Theorem 2]{silva2006asymptotic} and the Delta method.

\subsection{Conditional least squares}

The conditional least squares (CLS) method for estimation of parameters of GINAR($p$) processes was proposed by  \citet{klimko1978conditional}. We first define the modified parameter space which leaves out $\sigma^2_{\epsilon}$ from $\boldsymbol{\theta}$. Let $\boldsymbol{\eta} = (\alpha_1, \alpha_2, \ldots, \mu_{\epsilon})^T$. Then define
\begin{equation}
    U_n(\boldsymbol{\eta}) = \sum_{t=p+1}^n \{X_t - \mu_{X_t|X_{t-1},...,X_{t-p}}\}^2.
\end{equation}
The CLS estimator, $\widehat{\boldsymbol{\eta}}_{CLS}$, satisfies
\begin{align}
\label{cls.min}
    \boldsymbol{\widehat{\eta}}_{CLS} = \underset{\boldsymbol{\eta} \in \Theta}{\arg \min} \ U_N(\boldsymbol{\eta}).
\end{align}

The following is due to \citet[][Proposition 6.1]{latour1998existence}.

\begin{thm}
\label{cls.asymptotics}
    Let $\{X_t:t \in \mathbb{Z}\}$ be a GINAR($p$) process such that $E(\epsilon_t^3) < \infty$ for all $t \in \mathbb{Z}$.
Let $\{Y_k^{(j)}\}$ be as Definition \ref{gen.thin.def} where $j$ indicates that it is associated with the $j^{th}$ thinning operator $\alpha_j \odot$, and $E\{(Y_k^{(j)})^3)\} < \infty$, for all $j=1,\ldots,p$ and $k \geq 1$. Then the CLS estimator, $\widehat{\boldsymbol{\eta}}_{CLS}$, has the following asymptotic distribution
    \begin{align*}
        \sqrt{n}\left(\boldsymbol{\widehat{\eta}_{CLS}}-\boldsymbol{\theta}\right) \xrightarrow{\text{d}} N\left(\boldsymbol{0}, \boldsymbol{\Sigma}_{CLS}\right)
    \end{align*}
    as $n \rightarrow \infty$, where $\boldsymbol{\Sigma}_{CLS} = \mathbf{V^{-1}WV}$. In the definition of $\boldsymbol{\Sigma}_{CLS}$,
    \begin{align*}
        \mathbf{V} &= \begin{bmatrix}
\boldsymbol{\Gamma} + \mu_X^2 \boldsymbol{1}_p \boldsymbol{1}_p^T & \mu_X\boldsymbol{1}_p \\
\mu_X\boldsymbol{1}_p^T & 1 
\end{bmatrix},
\end{align*}
and 
\begin{align*}
\boldsymbol{V}^{-1} &= \begin{bmatrix}
    \boldsymbol{\Gamma}^{-1} & -\mu_X \boldsymbol{\Gamma}^{-1}\boldsymbol{1}_p \\
    -\mu_X \boldsymbol{1}_p^T \boldsymbol{\Gamma}^{-1} & 1+\mu_X^2 \boldsymbol{1}_p^T \boldsymbol{\Gamma}^{-1} \boldsymbol{1}_p
\end{bmatrix}, 
\end{align*}
where $\boldsymbol{1}_p$ is a $p \times 1$ matrix of 1's.
    The elements of the $\textbf{W}$ matrix are
    \begin{align*}
        W_{lj} = \sum_{k=1}^p \beta_k E\left(X_{p-l+1}X_{p-j+1}X_{p-k+1}\right) + \sigma^2_{\epsilon} V_{lj},  \quad 1 \leq l, j \leq p.
    \end{align*}
\end{thm}
This method is computationally more tractable than CML, however it does not take into account other conditional moment restrictions, like the conditional heteroscedasticity, leading to a possible loss of efficiency in comparison to CML estimation \citep{hadri2009maximum}. 

We can estimate the innovation variance $\sigma^2_{\epsilon}$ using a two-step CLS 
method~\citep{karlsen1988consistent,ristic2012estimation,bourguignon2019extended} -- we minimize
\begin{align*}
    S_n(\boldsymbol{\eta}) = \sum_{t=p+1}^n \left[\{X_t - \mu_{X_t|X_{t-1},...,X_{t-p}}\}^2 - \sigma^2_{X_t|X_{t-1},\ldots, X_{t-p}}\}\right]^2,
\end{align*}
with respect to $\sigma^2_{\epsilon}$, while replacing $\boldsymbol{\alpha}$ and $\mu_{\epsilon}$ with the CLS estimates from (\ref{cls.min}).
(Also see \citet{freeland2010true}, \citet{bourguignon2015first}, and \citet{huang2021new}.)

\subsection{Pseudo maximum likelihood}

For the pseudo maximum likelihood method we approximate the transition probability using normal distribution with mean and variance equal to the conditional mean and variance presented in Lemma \ref{stat.properties}. Let
\begin{align*}
    \ell_P(\boldsymbol{\theta}) &=
     - \frac{1}{2} \sum_{t=p+1}^n \log\left( 2\pi\sigma_{X_t|X_{t-1},\ldots, X_{t-p}}^2(\boldsymbol{\theta}) \right)  + \sum_{t=p+1}^n \frac{ \left(x_t - \mu_{X_t|X_{t-1},\ldots, X_{t-p}}(\boldsymbol{\theta})\right)^2}{2\sigma_{X_t|X_{t-1},\ldots, X_{t-p}}^2(\boldsymbol{\theta})}.
\end{align*}
Then the pseudo maximum likelihood estimator $\widehat{\boldsymbol{\theta}}_p$ is then
\begin{align*}
    \widehat{\boldsymbol{\theta}}_{P} = \underset{\boldsymbol{\theta} \in \Theta}{\arg \max} \ \ell_P(\boldsymbol{\theta}).
\end{align*}

\begin{thm}
\label{thm:pseudo.asymptotics}
 Let $\{X_t:t \in \mathbb{Z}\}$ be a GINAR($p$) process. The pseudo maximum likelihood estimator, $\widehat{\boldsymbol{\theta}}_{P}$, has the asymptotic distribution
\begin{equation}
    \sqrt{n}\left(\widehat{\boldsymbol{\theta}}_{P}-\boldsymbol{\theta}\right) \xrightarrow{\text{d}} N\left(\boldsymbol{0}, \boldsymbol{\Sigma}_{P}\right),
\end{equation}
 as $n \rightarrow \infty$, where 
 \begin{align*}
 \boldsymbol{\Sigma}_{P} &= \mathbf{{U(\boldsymbol{\theta})}^{-1}Z(\boldsymbol{\theta}){U(\boldsymbol{\theta})}^{-1}},\\
     \mathbf{Z(\theta)} &= \underset{n \rightarrow \infty}{lim} \ \frac{1}{n} E\!\left( \frac{\partial \ell_P(\boldsymbol{\theta})}{ \partial \boldsymbol{\theta}} \left({\frac{\partial \ell_P(\boldsymbol{\theta})}{ \partial \boldsymbol{\theta}}}\right)^T   \right),
\end{align*}
and
\begin{align*}
     \mathbf{U(\theta)} &= \underset{n \rightarrow \infty}{lim} \ \frac{1}{n} E\!\left(\frac{\partial^2 \ell_P(\boldsymbol(\theta))}{\partial \boldsymbol{\theta} \partial \boldsymbol{\theta}^T} \right).
 \end{align*}
\end{thm}

This method is computationally much more tractable compared to CML, and also has a relatively simple implementation.

\subsection{Saddlepoint methods}

Saddlepoint approximation techniques were introduced by \citet{daniels1954}, and are used in a wide range of applications. An advantage of these methods is that they tend to provide accurate approximations even for small sample sizes, and is often believed to be more accurate than using a normal approximation ~\citep[e.g.,][]{reidsaddle, davison2003statistical}. \citet{pedeli2015likelihood} propose the saddlepoint approximation method for the GINAR($p$) case with binomial thinning. Their main idea is that the convolutions in the likelihood expression can be removed by considering the corresponding moment-generating functions. First, they note that the conditional cumulant generating function for the GINAR($p$) process with binomial thinning and a general innovation distribution is
\begin{align*}
    K_t(u) &= \log E[\exp(uX_t)|X_{t-1},...,X_{t-p}] \\
    &= \sum_{j=1}^{p} x_{t-j} \log(1-\alpha_j + \alpha_j \exp(u)) + K_{\epsilon_t}(u).
\end{align*}
Generalizing to other thinning operators we get
\begin{align*}
    K_t(u) &= 
 \sum_{j=1}^{p} K_{\alpha \odot X_{t-j}}(u) + K_{\epsilon_t}(u).
\end{align*}

Then, an approximation to the true conditional log likelihood is provided by the saddlepoint approximation is
\begin{align*}
    \ell_S(\boldsymbol{\theta}) = \sum_{t=p+1}^n \left(\frac{1}{2\pi K^{''}_t(\Tilde{u_t})} \right)^{1/2} \exp\left(K_t(\Tilde{u_t})-\Tilde{u_t}x_t\right),
\end{align*}
where $\Tilde{u_t}$ is found by solving $K'_t(u) = x_t$, and the derivatives are taken with respect to $u$. \citet{pedeli2015likelihood} propose using a quasi-likelihood based approach to show that the resulting estimator is asymptotically normal for the binomial thinning case, however they remark that the estimator is not consistent. They posit that the saddlepoint method is more computationally efficient than the CML method with transition probabilities calculated using the inefficient equation in Theorem \ref{thm:trans.prob}, especially for cases where observed counts are large.

\subsection{Spectral-based Whittle estimation}
We can also consider estimating the parameters on the basis of a spectral analysis of the process. The Whittle criterion is
\begin{equation}
    \ell_W(\boldsymbol{\theta}) = \frac{1}{N} \sum_{j=1}^{[N/2]} \left(\log f(\nu_j, \boldsymbol{\theta}) + \frac{I_N(\nu_j)}{f(\nu_j, \boldsymbol{\theta})}\right),
\end{equation}
where $f(\nu_j, \boldsymbol{\theta})$ is the spectral density function and $I_N(\nu_j)$ is the periodogram at frequency $\nu_j = 2\pi j/N$ \citep{silva2005difference}. 
The Whittle estimate of $\boldsymbol{\theta}$, $\boldsymbol{\widehat{\theta}_{W}}$, is
\begin{align*}
    \widehat{\boldsymbol{\theta}}_{W} = \underset{\boldsymbol{\eta} \in \Theta}{\arg \min} \ \ell_W(\boldsymbol{\theta}).
\end{align*}
The Whittle estimates are asymptotically normal (e.g, \citet{rice1979estimation}). However, the asymptotic variance depends on the fourth order spectra of the process, which is difficult to obtain for the case of a GINAR process.

For a large class of processes the Whittle likelihood is known to provide biased estimates. Hence,  \citet{sykulski2019debiased} propose the debiased Whittle likelihood which replaces the spectral density with the expected value of the periodogram while still being computationally efficient with $O(n\log n)$ efficiency. The authors show that employing this method leads to less biased estimates compared to using the Whittle likelihood.

\subsection{Other methods}

Other estimation methods have also been proposed. For instance, the squared difference (SD) estimator is proposed for the GINAR(1) process with binomial thinning and Poisson innovations have, which provides bias reduced estimates using a method of moments approach \citep{bourguignon2015improved}. 
Bootstrap based techniques have also been studied. For instance, \citet{jentsch2019bootstrapping} and \citet{weiss2019bootstrap} propose some flexible bootstrap techniques for the GINAR($p$) process with binomial thinning which could be extended to a generalized thinning case. \citet{lu2022new} propose a new parameter estimator based on empirical likelihood  
for the GINAR(1) process with binomial thinning and Poisson innovations which has a small bias and is efficient as shown by simulation studies. Generalized method of moments have also been used for estimation for different models under the GINAR($p$) class, some examples are \cite{brannas1993estimation} and \citet{zhang2010inference}.

\section{Statistical inference and forecasting}
\label{sec:Inference}

In this section we discuss methods of inference for GINAR($p$) processes including building confidence regions, forecasting, and model selection.

\subsection{Confidence regions}

We can calculate confidence intervals for the model parameters based on the asymptotic theory given in Section \ref{sec:est.methods}. An estimate of the asymptotic variance for methods such as CML, CLS, Yule-Walker and pseudo maximum likelihood can be obtained by substituting the appropriate estimate into the expression for the asymptotic covariance matrix. We can also estimate the covariance matrix using the Hessian, calculated numerically from the optimization algorithm that generates the estimator.

For example, with an estimated covariance matrix for the parameters of $\boldsymbol{\Sigma}$, an approximate $(1-\nu) 100\%$ confidence interval for $\theta_j$ is 
\begin{align*}
    \widehat{\theta}_{\text{j}} \pm z_{1-\nu/2} \left[\boldsymbol{\Sigma}\right]_{\text{jj}}
\end{align*}
and a $100(1-\nu)$\% simultaneous confidence region for $\boldsymbol{\theta}$ is
\begin{align*}
    \left\{\boldsymbol{\theta}:(\widehat{\boldsymbol{\theta}} - \boldsymbol{\theta}) \boldsymbol{\Sigma}^{-1} (\widehat{\boldsymbol{\theta}} - \boldsymbol{\theta})^T \leq \frac{\chi^2_{p, 1-\nu}}{n}  \right\}.
\end{align*}
\citet{weiss2011simultaneous} use simulations to compare the confidence interval coverage for the CLS and CML methods for a GINAR(1) process with binomial thinning and Poisson innovations. 
They indicate that the CML method has better coverage properties.  The CLS method has lower coverage when the dependence parameter $\alpha_1$ is large for shorter time series.

\subsection{Forecasting}
\label{sec:forecasting}

When it comes to forecasting GINAR($p$) processes, we ideally want a method that produces integer-valued forecasts. \citet{duli1991} propose the minimum variance predictor of $X_{n+1}$, denoted $\widehat{X}_n(1)$, to be the conditional expectation. Similarly then, the \textit{k}-step ahead prediction is given by
\begin{align*}
    \widehat{X}_n(k) &= \mu_{X_{n+k}|X_1,...,X_n} = \sum_{j=1}^p E(\alpha_j \odot X_{n+k-j}) + \mu_\epsilon =  \sum_{j=1}^p \alpha_j \widehat{X}_n(k-j) + \mu_\epsilon.
\end{align*}
Although this method produces forecasts with minimum mean square errors, it this has the disadvantage of producing non-integer forecasts due to the multiplications involved. Hence, other methods have been proposed in the literature. \citet{freeland2004forecasting}, present the \textit{k}-step ahead predictive distribution for the GINAR(1) process with binomial thinning and Poisson innovations, but analytical solutions are not easily derived for higher order lags and other innovation distributions. They provide the asymptotic distribution of the \textit{k}-step ahead probability mass function which can be used to construct confidence intervals for the probability associated with any value in the forecast distribution. \citet{bu2008model} propose a forecasting methodology for the GINAR($p$) process with binomial thinning and Poisson innovations,
using the transition probability function of the process. They treat the model as a Markov chain with a maximum count of $M$, and obtain forecasts using the resulting transition matrix. They also posit a methodology for calculating confidence intervals for these forecast probabilities.  

Other commonly used forecasting approach uses bootstrap based prediction methods, which allow for a distribution free approach, making them more flexible to incorporate higher order lags, and different thinning and innovation distributions. \citet{cardinal1999application}, \citet{kim2008non}, \citet{bisaglia2019model}, implement bootstrap based prediction methods. \citet{bisaglia2019model} also carries out an extensive simulation study comparing their proposed method to that of \citet{cardinal1999application} and  \citet{kim2008non}. 

For the purpose of the application presented in this paper, the forecasting distribution is calculated by means of a Monte Carlo procedure. We use the median of this distribution as a point estimate and prediction interval can be obtained from the quantiles. This ensures that we always obtain an integer-valued forecast. The steps for calculating the \textit{h}-step ahead forecast distribution, denoted $X_{n+h}$, for a GINAR($p$) process is as follows:

\begin{itemize} 
  \item[1.] Generate one realization of $\widehat{\alpha}_{j}\odot X^b_{n+h-j} = \sum_{j=1}^{X^b_{n+h-j}} Y_k^j$ where $\{Y_k^j\}$ are IID RVs as described in Definition \ref{gen.thin.def}, for $j=1,2,\ldots,p$;
  \item[2.] Generate one realization of the innovation distribution $\{\epsilon^b_{n+h}\}$ which are IID RVs from a chosen non-negative distribution with parameters $\widehat{\mu}_{{\epsilon}}$ and $\widehat{\sigma}^2_{\epsilon}$;
  \item[3.] Set $X^b_{N+h} = \sum_{j=1}^{p}\widehat{\alpha}_{j}\odot X^b_{n+h-j} + \epsilon^b_{n+h}$;
  \item[4.] Repeat steps 1--3, $b$ times, for large $b$.
\end{itemize}

\subsection{Model selection}

\citet{weiss2019performance} carry out a comprehensive simulation study on the performance of AIC and BIC for model selection in the context of count time series models. The authors choose candidate models to provide insight into tasks such as identifying serial correlation, overdispersion or zero-inflation, order selection and non-nested model families. Their findings show that both AIC and BIC are generally successful model selection criteria -- for shorter series, AIC is recommended, while BIC outperforms AIC for longer series.

\citet{alzahrani2018model} discuss other methods for model selection which are useful when the competing models are non-nested, which is usually the case. They develop an effective algorithm in the Bayesian framework, implementing direct computation of the marginal likelihood and using the deviance information criteria (DIC). 
\citet{bu2008model} present a model selection approach using a new residual process which is defined for each of the $p$+1 model components (the $p$ thinning operations of past values and the innovation sequence). Here they look at the residuals of each model component, in addition to residuals of the fitted model. They posit that this allows for a more robust analysis into the goodness of fit of the model. \citet{freeland2004analysis} also suggest defining two sets of residuals for the GINAR(1) process with binomial thinning, one for each random component part of the model. However, the challenge here is interpreting these residuals when the lags are higher. \citet{park2012diagnostic} present two new expected residual calculations for the GINAR(\textit{p}) process with binomial thinning, and demonstrate its usage in assessing model adequacy and predictive performance. They show that these residuals can be useful in model selection, identifying overdispersion and correlation arising from incorrect parameter estimates or choice of model order $p$. These residuals require that the conditional probabilities, $P(X_t|X_{t-1}, \dots, X_{t-p})$ and $P(\epsilon_t|X_{t-1}, \dots, X_{t-p})$ are specified, hence this method can be extended to the GINAR($p$) case whenever the conditional distributions are available. 
\citet{forughi2022portmanteau} develops some portmanteau test statistics to check goodness of fit for GINAR($p$) processes and derive asymptotic distributions of the test statistics.

\section{Simulation studies}
\label{sec:sim.studies}

Up to this point we have reviewed and discussed a number of different estimation, inference, and forecasting methods that can be applied to the statistical analysis of GINAR($p$) processes.  Now we compare the performance of different estimation methods for finite samples, taking time to compare the performance as we vary the order $p$, as well as the thinning operator and innovation process used to define the GINAR process.

Throughout, Po-INAR refers to a GINAR process with binomial thinning and Poisson innovations, NB-INAR refers to binomial thinning and negative binomial innovations, and Geom-INAR refers to negative binomial thinning defined with geometric random variables along with Poisson innovations.  In all simulations we estimate each quantity of interest (the bias, standard deviation (SD), root mean squared error (RMSE), or coverage) using 10,000 replicates, and estimate standard errors for each quantity using 10,000 bootstrap samples.

\label{sec:Simulation Study}

\subsection{Comparing estimation methods}

In our first simulation we estimate the parameters of the most commonly used Po-INAR(1) process when $\alpha_1 = 0.5$ and $\mu_{\epsilon} = 1$.  Table~\ref{tab.inar1.sims.poinar} shows estimates of the bias, SD, and RMSE for both parameters of the process using different estimation methods as we vary the sample length $n$, with $n=100$, $500$ and 1,000.  For most estimation methods, we learn that the bias tends to get closer to zero as we increase $n$, and that naturally the SD and RMSE gets smaller as $n$ is increased.  However, due to the lack of consistency discussed earlier, the bias of the saddlepoint estimator does not go to zero.  For longer sample lengths the CML methods tends to do best in terms of bias, SD, and RMSE, although the pseudo-likelihood method is comparable in performance.  There is a loss of efficiency for other estimators.  For the CML method we found no difference in the results when using the slow transition probability calculation given by Theorem~\ref{thm:trans.prob} and the much faster chf-based calculation in Proposition~\ref{prop:davies.tp}, evaluating the integrals using Gauss quadrature with 300 weights.

\begin{table}[t]
\caption{\label{tab.inar1.sims.poinar}Estimated bias, SD, and RMSE when estimating the parameters of a Po-INAR(1) process with $\alpha_1=0.5$ and $\mu_{\epsilon}=1$. The maximum standard error for each quantity is 0.002 for estimating $\alpha_1$ and 0.004 for estimating $\mu_{\epsilon}$.}
\vspace{.5cm}
\centering
\renewcommand{\arraystretch}{0.7}
\hspace*{-1.3cm}
\begin{tabular}{ll|rr|rr|rr} 
& & \multicolumn{2}{c}{$n=$ 100} & \multicolumn{2}{c}{$n=$ 500} & \multicolumn{2}{c}{$n=$1,000} \\
Method & &  $\widehat{\alpha}_1$ & $\widehat{\mu}_{\epsilon}$ & $\widehat{\alpha}_1$ & $\widehat{\mu}_{\epsilon}$ & $\widehat{\alpha}_1$ & $\widehat{\mu}_{\epsilon}$ \\ 
\hline
{CML}  & Bias & -0.009 & 0.013 & -0.002 & 0.003 & -0.001 & 0.001\\
& SD   & 0.076 & 0.166 & 0.032 & 0.073 & 0.023 & 0.051\\
& RMSE & 0.076 & 0.166 & 0.033 & 0.073 & 0.023 & 0.051\\
\hline 
{CLS}  & Bias & -0.026 & 0.049 & -0.006 & 0.011 & -0.002 & 0.004\\
& SD   & 0.093 & 0.203 & 0.042 & 0.089 & 0.030 & 0.063\\
& RMSE & 0.097 & 0.209 & 0.043 & 0.090 & 0.030 & 0.063\\
\hline 
{YW}  & Bias & -0.032 & 0.059 & -0.006 & 0.012 & -0.003 & 0.005\\
& SD   & 0.093 & 0.206 & 0.042 & 0.089 & 0.029 & 0.063\\
& RMSE & 0.099 & 0.215 & 0.042 & 0.090 & 0.029 & 0.063\\
\hline 
{Pseudo}  & Bias & -0.011 & 0.008 & -0.002 & 0.002 & -0.001 & 0.001\\
& SD   & 0.083 & 0.176 & 0.035 & 0.077 & 0.025 & 0.055\\
& RMSE & 0.083 & 0.176 & 0.035 & 0.077 & 0.025 & 0.055\\
\hline 
{Whittle}  & Bias & -0.017 & 0.012 & -0.003 & 0.003 & -0.002 & 0.000\\
& SD   & 0.094 & 0.187 & 0.042 & 0.083 & 0.030 & 0.058\\
& RMSE & 0.096 & 0.187 & 0.042 & 0.083 & 0.030 & 0.058\\
\hline 
{Saddle}  & Bias & -0.045 & 0.086 & -0.037 & 0.073 & -0.036 & 0.071\\
& SD   & 0.075 & 0.169 & 0.033 & 0.073 & 0.023 & 0.052\\
& RMSE & 0.087 & 0.189 & 0.049 & 0.103 & 0.043 & 0.088\\

\end{tabular}
\end{table}

In Table~\ref{tab.inar1.sims.nbinar} we move to using a NB-INAR(1) process instead of the Po-INAR(1) process.   Both processes are defined with binomial thinning but we now use negative binomial innovations with mean $\mu_{\epsilon}$ and variance $\sigma^2_{\epsilon} = \mu_{\epsilon} + r\mu_{\epsilon}^2$ with $r=1$ for the NB-INAR(1) process, generating more overdispersion as compared to Po-INAR(1) process. In general the results stay consistent with the results to the Po-INAR(1) simulation. However, the pseudo-likelihood estimator does worse in comparison to CML and the saddlepoint estimator also performs much worse. Also, it is much harder to estimate $r$ compared to the other parameters of the model. One can clearly see the merits of using CML here as it tends to do best in terms of bias, SD and RMSE.  

\begin{table}[t]
\caption{\label{tab.inar1.sims.nbinar}Estimated bias, SD, and RMSE when estimating the parameters of a NB-INAR(1) process with $\alpha_1=0.5$, $r=1$ and $\mu_{\epsilon}=1$. The maximum standard error for each quantity is 0.006 for estimating $\alpha_1$, 0.05 for estimating $r$, and 0.05 for estimating $\mu_{\epsilon}$.}
\vspace{.5cm}
\centering
\renewcommand{\arraystretch}{0.7}
\begin{tabular}{ll|rrr|rrr|rrr} 
& & \multicolumn{3}{c}{$n=$ 100} & \multicolumn{3}{c}{$n=$ 500} & \multicolumn{3}{c}{$n=$ 1,000} \\
Method & &  $\widehat{\alpha}_1$  & $\widehat{\mu}_{\epsilon}$  & $\widehat{r}$ & $\widehat{\alpha}_1$ & $\widehat{\mu}_{\epsilon}$ & $\widehat{r}$ & $\widehat{\alpha}_1$ & $\widehat{\mu}_{\epsilon}$ & $\widehat{r}$\\ 
\hline
{CML}\hspace{-1cm}  & Bias & -0.013 & 0.021 & -0.014 & -0.002 & 0.003 & -0.005 & -0.002 & 0.002 & -0.001\\
& SD   & 0.068 & 0.184 & 0.505 & 0.028 & 0.079 & 0.216 & 0.020 & 0.055 & 0.151\\
& RMSE & 0.069 & 0.186 & 0.505 & 0.029 & 0.079 & 0.216 & 0.020 & 0.055 & 0.151\\
\hline {CLS}  & Bias & -0.030 & 0.054 & -0.071 & -0.006 & 0.011 & -0.013 & -0.003 & 0.004 & -0.006\\
& SD   & 0.095 & 0.233 & 0.713 & 0.043 & 0.101 & 0.315 & 0.030 & 0.072 & 0.225\\
& RMSE & 0.099 & 0.239 & 0.716 & 0.043 & 0.102 & 0.315 & 0.031 & 0.072 & 0.225\\
\hline {YW}  & Bias & -0.031 & 0.057 & -0.058 & -0.006 & 0.011 & -0.010 & -0.003 & 0.008 & -0.004\\
& SD   & 0.094 & 0.233 & 0.705 & 0.042 & 0.103 & 0.316 & 0.030 & 0.073 & 0.224\\
& RMSE & 0.098 & 0.240 & 0.708 & 0.043 & 0.103 & 0.316 & 0.031 & 0.073 & 0.224\\
\hline {Pseudo}  & Bias & -0.029 & 0.053 & -0.084 & -0.006 & 0.011 & -0.024 & -0.003 & 0.005 & -0.009\\
& SD   & 0.098 & 0.240 & 0.578 & 0.043 & 0.102 & 0.267 & 0.030 & 0.071 & 0.190\\
& RMSE & 0.103 & 0.245 & 0.584 & 0.043 & 0.102 & 0.268 & 0.030 & 0.071 & 0.190\\
\hline {Whittle}  & Bias & -0.013 & 0.053 & -0.040 & -0.001 & 0.008 & 0.002 & 0.000 & 0.000 & 0.016\\
& SD   & 0.094 & 0.227 & 0.690 & 0.043 & 0.101 & 0.318 & 0.030 & 0.070 & 0.217\\
& RMSE & 0.095 & 0.233 & 0.691 & 0.043 & 0.102 & 0.318 & 0.030 & 0.069 & 0.217\\
\hline {Saddle}  & Bias & -0.075 & 0.147 & -0.490 & -0.064 & 0.127 & -0.468 & -0.063 & 0.123 & -0.466\\
& SD   & 0.074 & 0.208 & 0.318 & 0.034 & 0.146 & 0.142 & 0.028 & 0.066 & 0.099\\
& RMSE & 0.105 & 0.254 & 0.584 & 0.072 & 0.193 & 0.489 & 0.069 & 0.140 & 0.477\\
\end{tabular}
\end{table}

Table~\ref{tab.inar1.sims.geominar} compares the performance of different estimators when we simulate Geom-INAR(1) processes with $\alpha_1 = 0.5$ and $\mu_{\epsilon} = 1$, at the same sample lengths $n$.  For this process we have negative binomial thinning and Poisson innovations, leading to overdispersion that varies over time.   Also for the Geom-INAR(1) case the saddlepoint method is not available, and thus is not shown. The simulation results in Table \ref{tab.inar1.sims.geominar} show that CML and pseudo-likelihood are comparable in terms of Bias, SD and RMSE. We see that Whittle likelihood estimates perform much worse as compared to the Po-INAR(1) simulation, especially for smaller sample lengths with the RMSE for $\mu_{\epsilon}$ being almost two times larger than that of CML.

\begin{table}[t]
\caption{\label{tab.inar1.sims.geominar}Estimated bias, SD, and RMSE when estimating the parameters of a Geom-INAR(1) process with $\alpha_1=0.5$ and $\mu_{\epsilon}=1$. The maximum standard error for each quantity is 0.002 for estimating $\alpha_1$ and 0.007 for estimating $\mu_{\epsilon}$.}
\vspace{.5cm}
\centering
\renewcommand{\arraystretch}{0.7}
\hspace*{-1.3cm}
\begin{tabular}{ll|rr|rr|rr} 
& & \multicolumn{2}{c}{$n=$ 100} & \multicolumn{2}{c}{$n=$ 500} & \multicolumn{2}{c}{$n=$ 1,000} \\
Method & &  $\widehat{\alpha}_1$ & $\widehat{\mu}_{\epsilon}$ & $\widehat{\alpha}_1$ & $\widehat{\mu}_{\epsilon}$ & $\widehat{\alpha}_1$ & $\widehat{\mu}_{\epsilon}$ \\ 
\hline
{CML}  & Bias & -0.027 & 0.041 & -0.006 & 0.010 & -0.003 & 0.004\\
& SD   & 0.096 & 0.182 & 0.042 & 0.079 & 0.029 & 0.055\\
& RMSE & 0.100 & 0.186 & 0.042 & 0.080 & 0.029 & 0.055\\
\hline 
{CLS}  & Bias & -0.040 & 0.066 & -0.009 & 0.014 & -0.004 & 0.007\\
& SD   & 0.109 & 0.212 & 0.051 & 0.098 & 0.036 & 0.069\\
& RMSE & 0.116 & 0.222 & 0.052 & 0.099 & 0.037 & 0.069\\
\hline 
{YW}  & Bias & -0.042 & 0.073 & -0.009 & 0.015 & -0.004 & 0.008\\
& SD   & 0.106 & 0.209 & 0.051 & 0.097 & 0.036 & 0.069\\
& RMSE & 0.114 & 0.221 & 0.051 & 0.098 & 0.036 & 0.070\\
\hline 
{Pseudo}  & Bias & -0.022 & 0.022 & -0.005 & 0.006 & -0.002 & 0.001\\
& SD   & 0.106 & 0.210 & 0.045 & 0.091 & 0.032 & 0.063\\
& RMSE & 0.108 & 0.211 & 0.046 & 0.091 & 0.032 & 0.063\\
\hline 
{Whittle}  & Bias & -0.029 & 0.060 & -0.006 & 0.014 & -0.003 & 0.006\\
& SD   & 0.109 & 0.338 & 0.050 & 0.151 & 0.036 & 0.107\\
& RMSE & 0.113 & 0.344 & 0.051 & 0.152 & 0.036 & 0.107\\
\end{tabular}
\end{table}

We next consider estimation of GINAR processes when the order of the process, $p$, is higher, with $p=4$.  In this case we estimate the parameters of Po-INAR(4) and Geom-INAR(4) processes and $\vv{\alpha} = (0.1, 0.2, 0.1, 0.3)^T$ with $\mu_{\epsilon}=10$ when the sample length is $n=500$.  Again, the saddlepoint method is not available in this case.  Table~\ref{tab.inar4.sims.1.poinar} shows how the remaining estimators of the parameters for these two processes compare.  For all methods, estimating the thinning probability parameters $\vv{\alpha}$ tend to be easier than estimating the innovation mean $\mu_{\epsilon}$.  Across both processes, the CML method tends to perform the best with respect to bias, SD, and RMSE.  For the Po-INAR(4) process the pseudo likelihood and Whittle methods do comparably well as compared to the CML method, whereas for the Geom-INAR(4) process the Yule-Walker and pseudo methods are comparable to CML method. 

In the supplement we present additional simulations for the processes mentioned above. For the Po-INAR(1), NB-INAR(1) and Geom-INAR(1) processes we vary the innovation mean and dependence parameter. When varying the innovation mean, we see that the saddlepoint estimator improves in terms of RMSE for larger values of $\mu_{\epsilon}$.  
However, it has much worse bias properties than all other estimation methods for smaller values of $\mu_{\epsilon}$, for both NB-INAR(1) and Po-INAR(1) processes. A large bias and small RMSE suggests that saddlepoint estimator is consistently producing estimates that are far off from the true values, which is not desirable. Varying the values of the dependence parameter shows that CML has the lowest RMSE for all parameters estimated for all three processes and most methods have large bias, SD and RMSE as $\alpha_1$ gets close to the boundary point of $\alpha_1 = 1$. We also present simulations for the GINAR(4) processes discussed for varying $n$ which further supports the conclusion from this section which favor the CML estimation method. 

\begin{table}[t]
\caption{\label{tab.inar4.sims.1.poinar}Estimated bias, SD, and RMSE when estimating the parameters of a Po-INAR(4) process with $\alpha_1=0.1$, $\alpha_2 = 0.2$, $\alpha_3 = 0.3$ and $\alpha_4 = 0.1$ and $\mu_{\epsilon}=10$. The maximum standard error for each quantity is 0.002 for estimating $\alpha_1$, 0.003 for estimating $\alpha_2$, 0.001 for estimating $\alpha_3$, 0.006 for estimating $\alpha_4$, and 0.102 for estimating $\mu_{\epsilon}$. }
\vspace{.5cm}
\centering
\renewcommand{\arraystretch}{0.7}
\hspace*{-1.3cm}
\begin{tabular}{ll|rrrrr} 
& & \multicolumn{5}{c}{$n=$ 500}\\
Method & &  $\widehat{\alpha}_1$ & $\widehat{\alpha}_2$ & $\widehat{\alpha}_3$ & $\widehat{\alpha}_4$ & $\widehat{\mu}_{\epsilon}$ \\ 
\hline
{CML}  & Bias & 0.002 & -0.004 & -0.001 & -0.008 & 0.287 \\
& SD   & 0.043 & 0.043 & 0.042 & 0.044 & 1.830 \\
& RMSE & -0.004 & 0.043 & 0.042 & 0.044 & 1.853 \\
\hline
{CLS}  & Bias & -0.022 & 0.006 & -0.023 & 0.082 & -1.440\\
& SD   & 0.062 & 0.121 & 0.061 & 0.217 & 5.136 \\
& RMSE & 0.006 & 0.122 & 0.066 & 0.232 & 5.334 \\
\hline
{YW}  & Bias & -0.001 & -0.008 & -0.005 & -0.010 & 0.774 \\
& SD   & 0.043 & 0.043 & 0.044 & 0.042 & 1.963 \\
& RMSE & -0.008 & 0.044 & 0.044 & 0.044 & 2.110 \\
\hline
{Pseduo}  & Bias & -0.003 & -0.006 & -0.003 & -0.005 & 0.526 \\
& SD   & 0.044 & 0.043 & 0.043 & 0.042 & 1.849 \\
& RMSE & -0.006 & 0.044 & 0.043 & 0.042 & 1.923 \\
\hline
{Whittle}  & Bias & 0.002 & -0.004 & -0.001 & -0.008 & 0.287\\
& SD   & 0.043 & 0.043 & 0.042 & 0.044 & 1.830\\
& RMSE & -0.004 & 0.043 & 0.042 & 0.044 & 1.853\\
\end{tabular}
\hspace*{-1cm}
 
\end{table}

\begin{table}[t]
\caption{\label{tab.inar4.sims.1.geom}Estimated bias, SD, and RMSE when estimating the parameters of a Geom-INAR(4) process with $\alpha_1=0.1$, $\alpha_2 = 0.2$, $\alpha_3 = 0.3$ and $\alpha_4 = 0.1$ and $\mu_{\epsilon}=10$. The maximum standard error for each quantity is 0.001 for estimating $\alpha_1$, 0.003 for estimating $\alpha_2$, 0.001 for estimating $\alpha_3$, 0.006 for estimating $\alpha_4$, and 0.102 for estimating $\mu_{\epsilon}$ }
\vspace{.5cm}
\centering
\renewcommand{\arraystretch}{0.7}
\hspace*{-1.3cm}
\begin{tabular}{ll|rrrrr} 
& & \multicolumn{5}{c}{$n=$ 500}\\
Method & &  $\widehat{\alpha}_1$ & $\widehat{\alpha}_2$ & $\widehat{\alpha}_3$ & $\widehat{\alpha}_4$ & $\widehat{\mu}_{\epsilon}$ \\ 
\hline
{CML}  & Bias & -0.002 & -0.007 & -0.003 & -0.009 & 0.700\\
& SD   & 0.043 & 0.043 & 0.042 & 0.042 & 1.928\\
& RMSE & 0.043 & 0.044 & 0.042 & 0.043 & 2.051\\
\hline
{CLS}  & Bias & -0.022 & 0.006 & -0.023 & 0.082 & -1.440\\
& SD   & 0.062 & 0.121 & 0.061 & 0.217 & 5.136\\
& RMSE & 0.066 & 0.122 & 0.066 & 0.232 & 5.334\\
\hline
{YW}  & Bias & -0.002 & -0.006 & -0.004 & -0.011 & 0.727\\
& SD   & 0.044 & 0.044 & 0.043 & 0.043 & 1.978\\
& RMSE & 0.044 & 0.044 & 0.043 & 0.045 & 2.108\\
\hline
{Pseudo}  & Bias & -0.002 & -0.007 & -0.003 & -0.010 & 0.687\\
& SD   & 0.043 & 0.044 & 0.043 & 0.042 & 1.944\\
& RMSE & 0.043 & 0.044 & 0.043 & 0.043 & 2.062\\
\hline
{Whittle}  & Bias & 0.001 & -0.004 & -0.001 & -0.008 & 0.365\\
& SD   & 0.044 & 0.043 & 0.043 & 0.044 & 2.247\\
& RMSE & 0.044 & 0.044 & 0.043 & 0.044 & 2.277\\

\end{tabular}
\hspace*{-1cm}
\end{table}

In conclusion, the main takeaway from these simulations is that CML performs well over all other methods, even when changing the model order $p$, the values of the parameters, thinning operation, and innovation distribution.    However the CML method is more computationally intensive.  We recommend the use of the Davies method with Gauss quadrature which is much faster than using the exact convolution given in Theorem~\ref{thm:trans.prob} to calculate the transition probabilities.
If further speed increases are needed, employing other comparable estimation methods like pseudo maximum likelihood could be advantageous, however compared to CML, the accuracy of other estimation methods are more variable as the thinning operator and innovation distribution is varied in the GINAR($p$) process.

\section{Application}
\label{sec:Application}

\begin{figure}[t]
    \centering
    \includegraphics[]{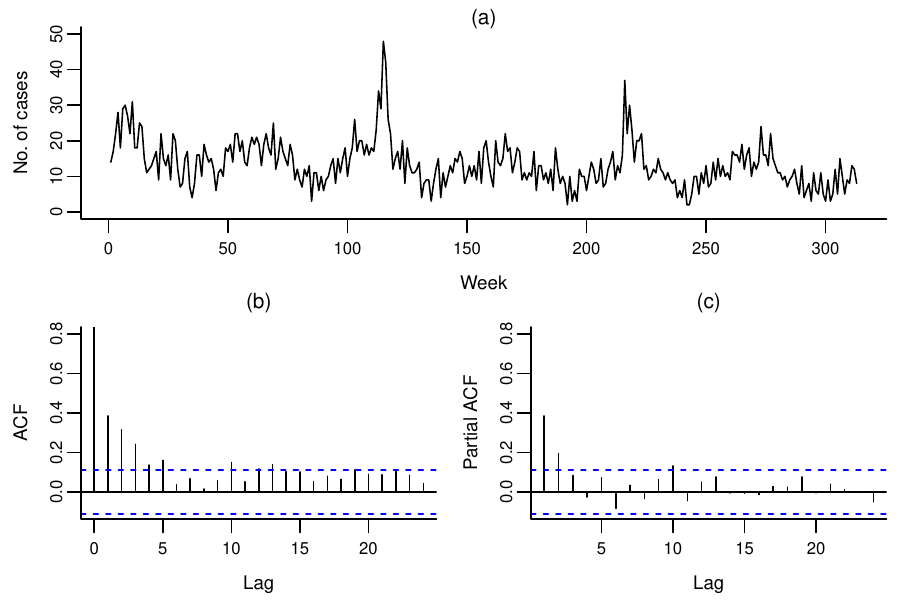}
    \caption{\label{fig:dataset}(a) Time series plot of the weekly meningococcal disease cases; (b) Sample ACF and (c) sample PACF of residuals after removing a yearly sinusoidal seasonality term.}
\end{figure}

Figure~\ref{fig:dataset} shows a summary of a time series of $n=313$ weekly meningococcal disease cases in Germany, over the period 2001--2006.  The data was studied by \citet{pedeli2015likelihood}.  Examining the time series plot in Figure~\ref{fig:dataset}(a) we see a strong yearly seasonal component, but also evidence of a mean-variance relationship with possible overdispersion over time.  Ignoring possible serial dependence, we remove the seasonal component by fitting a Poisson generalized linear model with a yearly sinusoidal term.  Figure~\ref{fig:dataset}(b) and (c) shows plots of the sample ACF and sample partial ACF up to lags 25 for the residuals, indicating that there is indeed time series dependence data after we account for this seasonal term.  Using standard diagnostics for autoregressive moving average  processes~\cite[e.g.][]{Brock} an autoregressive process of order 2 or 3 might be reasonable this time series.  However, autoregressive processes are a poor approximation for count series, and fail to account for possible overdispersion in the series.  Instead, we fit a number of different GINAR($p$) processes to the series, accounting for a time-varying seasonality using the following model for the time-varying log innovation mean:
\begin{align*}
    \log \mu_{\epsilon,t} = b_0 + b_1 \sin(2\pi t/52) + b_2 \cos(2\pi t/52).
\end{align*}

Table \ref{tab.disease.model} shows a summary of different GINAR($p$) processes fit to the count series, where we let $p=1, 2, 3,$ and $4$. As in the simulation study, we consider Po-INAR processes (binomial thinning with Poisson innovations), NB-INAR processes (binomial thinning with negative binomial innovations) and Geom-INAR processes (negative binomial thinning with Poisson innovations). We fit each process using the CML method, with the Davies method to calculate the transition probabilities.  For each process we tabulate the AIC, root mean squared error, one step ahead forecast coverage (using the Monte Carlo method described in Section \ref{sec:forecasting} and averaged over all time points) and the P-value of Ljung-Box test to examine whether or not the estimated innovations are a sample of IID noise (we use 20 lags for each test).

\begin{table}[t]
\caption{\label{tab.disease.model}A comparison of different model selection and diagnostic criteria for different GINAR processes fit to the meningococcal disease series.  All fitted processes contain a yearly seasonality component.  Values in bold indicate the processes for which the AIC and RMSE values are the smallest.}
\vspace{.5cm}
\centering
\begin{tabular}{l|cccc} 
GINAR process  & AIC & RMSE & Forecast coverage & Ljung-Box P-value \\ 
\hline
Po-INAR(1) & 1857.9 & 4.72 & 90.0\% & $\scriptstyle <$ 0.0001\\ 
NB-INAR(1) & 1814.9 & 4.68 & 96.5\% & $\scriptstyle <$ 0.0001\\
Geom-INAR(1) & 1825.2 & 4.65 & 93.3\% & 0.003\\
\hline
Po-INAR(2) & 1837.7 & 4.62 & 90.4\% & 0.019\\
NB-INAR(2) & 1800.1  & 4.77 & 95.5\% & 0.368 \\
Geom-INAR(2) & 1806.9 & 4.57 & 93.0\% & 0.331 \\
\hline
Po-INAR(3) & 1817.2	& 4.60	& 90.6\%	& 0.172\\
NB-INAR(3) & 1788.9	& 4.58 &	96.8\%	& 0.370 \\
Geom-INAR(3) & 1799.3	& 4.55	& 92.3\%	& 0.589 \\ 
\hline
Po-INAR(4) & 1818.5	& 4.58 & 88.3\%	& 0.130\\
NB-INAR(4) & \textbf{1782.0} & 4.56 & 96.8\%	& 0.226 \\
Geom-INAR(4) & 1792.4 &	\textbf{4.53} & 90.0\% & 0.455 \\ 
\end{tabular}
\end{table}

Table \ref{tab.disease.model} indicates that the NB-INAR(4) process is preferred with reference to minimizing the AIC and achieving good forecast coverage, whereas the Geom-INAR(4) process obtains a slightly smaller RMSE value.   On balance the NB-INAR(4) process is preferable.  All these processes give a better goodness of fit relative to using processes of order 3, as \citet{pedeli2015likelihood} considered -- it is worth noting however that they preferred the NB-INAR process as we do.
Considering GINAR($p$) processes with $p=5, 6, 7$ and $8$ did not greatly improve the fit and seemed to suggest we were overfitting to the data -- the confidence intervals for model parameters were too wide, as compared to using simpler processes.  In summary we learn that meningococcal disease cases in Germany from 2001--2006 exhibit strong dependencies over four-weekly (monthly) and yearly scales, and that there is significant overdispersion that should be accounted for.

\section{Conclusion and discussion}
\label{sec:Conclusion and Discussion}

In this article we provided a review of statistical inference techniques for analyzing generalized integer autoregressive processes. While these processes shares many characteristics with non-integer autoregressive processes, such as being a Markov process of a given order, through the selection of the thinning operator and innovation, we can accurately model positively valued count series with a variety of distributional and dispersion properties.   Also, forecasting methods can be designed to be exact.  While we focused most of our effort on determining the performance of estimating processes defined using binomial and geometric thinning, this could be expanded to the other thinning operators. (The supplement material provides some other examples of thinning operators.)

We extended the asymptotic results for different estimators of the process parameters to apply to GINAR processes, not just INAR processes defined with binomial thinning, as is commonplace in the literature.  Our simulations over different GINAR processes show that some methods are competitive to conditional maximum likelihood (CML) for some processes, however the overall best choice is CML especially for longer sample sizes. CML also tends to do better compared to other methods for overdispersed processes and when the processes is close to being nonstationary (i.e., when $\sum_j \alpha_j$ is close to one). Therefore, for applications where we have an overdispersed or highly dependent series we should be using CML to estimate the parameters.  When calculating the transition probabilities the Davies method using Gauss quadrature yields the same results as the convolution calculation and is computationally more efficient. Hence, we recommend using the Davies method for CML estimation of GINAR($p$) process parameters. It is important to note that even with this faster implementation, CML can be computationally costly as compared to other estimation methods, especially when the model order $p$ is large. In cases where computational complexity is an issue, methods like pseudo maximum likelihood can be used for certain processes like GINAR(1) processes with binomial thinning and Poisson innovations, as their performance is comparable to CML.  However, caution should be taken when varying the thinning operator and innovation process. Varying the type of  GINAR process, the Yule-Walker method used extensively in the literature may not be an ideal choice for estimation of model parameters.  Similarly the performance of other (non-CML) estimators are more mixed. In particular, saddlepoint methods have issues, and thus are less effective for estimating GINAR process parameters.  Note that in this article we did not consider Bayesian methods \cite[see e.g.,][]{mccabe2005bayesian,bisaglia2016bayesian,c2022bayesian}.

There are many ways that this work could be extended. Inspired by time series processes for non-count series, we can introduce dependence in the innovations or in the variance structure to obtain the INARMA and INGARCH processes, respectively \citep[see e.g.,][Chapters 3 and 4, for definitions]{weiss2018introduction}.  
Due to the extra latent structure that is introduced to define these processes, estimating parameters is more involved.   It would be interesting to learn how the different estimation and forecasting methods perform, especially with other thinning operators and more complicated model orders.  Similarly in the future we would like to also study multivariate integer time series processes defined by thinning operators.  For a review of these multivariate processes see, e.g., \citet[Chapter 8]{weiss2018introduction}.  For more general reviews of processes for count time series also see \citet{karlis2016} and \citet{davis2021count}.

\newpage

{\renewcommand{\baselinestretch}{1.1}\setlength{\bibsep}{2pt}
\small\bibliographystyle{abbrvnat}\bibliography{GINAR_statistical_inference}}

\end{document}


\section*{Supplemental material for:
    Statistical Inference for Generalized Integer Autoregressive Processes}
\vspace{-.25cm}

\textbf{Pashmeen Kaur}$^{1,3}$ and
\textbf{Peter F. Craigmile}$^{2,4}$

$^1$
Department of Statistics,
The Ohio State University,
Columbus, OH 43210, USA

$^2$
Department of Mathematics and Statistics,
Hunter College,
New York, NY 10065, USA

$^3$\verb_kaur.138@osu.edu_ \;
$^4$\verb_peter.craigmile@hunter.cuny.edu_ \;

\textit{Last updated \today}
\vspace{.25cm}

 In Section \ref{sec:other GINAR} we review thinning operators and innovation distributions used to define generalized integer autoregressive (GINAR) processes studied in the literature. In Section \ref{sec:proofs} we provide proofs for results in the paper and in Section \ref{sec:simulations} we present additional simulation results. 

\section{Additional Thinning Operators and Innovation Distributions}
\label{sec:other GINAR}

We review further thinning operators and innovation distributions used in the literature that align with the GINAR model framework. Here are two further thinning operators, and a comprehensive list of innovation distrbutions and thinning operations used are provided in Table \ref{tab:examples1} and \ref{tab:examples2}.

\textbf{$\rho$-Binomial}: In this thinning operation $\{Y_k: k \in \mathbb{N}\}$ is a set of independent $\rho$-Bernoulli RVs, a distribution introduced by \citet{kolev2000inflated}. A RV, $Y$, is said to have a $\rho$-Bernoulli distribution with success probability $\alpha \in [0,1]$ and dispersion parameter $\rho \in [0,1)$ if the probability mass function is given as 
\begin{align}
\label{eq:rhothin}
    P(Y=y) = \begin{cases}
        1 - \alpha  & \text{ if } y =0; \\
        \alpha \left(\frac{\rho}{1+\rho} \right)^{y-1} \left(\frac{1}{1+\rho} \right) & \text{ if } y=1, 2,\ldots.
    \end{cases}
\end{align}
We say that $Y \sim Bern(\alpha, \rho)$. Then the \textbf{$\rho$-Binomial thinning operator} is defined by using \eqref{eq:rhothin} as the counting sequence.

\textbf{Modified Negative Binomial Thinning}: As compared with negative binomial thinning, this thinning operator has slightly modified bounds given by
\begin{align*}
\alpha \odot X = \sum_{k=1}^{X+1} Y_k,
\end{align*}
where $Y_k \sim$ Geometric(1/(1+$\alpha$)). This modified definition was introduced to circumvent the problem of having constant zeroes over time since $\alpha \odot 0 = 0$ and is used in the definition of minifaction INAR(1) models which are not a GINAR process \citep{aleksic2021geometric}.

\begin{table}[p]
\caption{\label{tab:examples1} Examples of GINAR processes found in literature. 'NA' indicated that a closed form expression for the marginal distribution does not exist.}\centering
\begin{tabular}{|p{3cm}p{4cm}p{4cm}c p{3cm}|}
\hline
\textbf{Process} & \textbf{Thinning} & \textbf{Innovation} & \textbf{Marginal} & \textbf{Reference} \\
& \textbf{Operator} & \textbf{Dist.} & \textbf{Dist.}& \\
\hline
\textbf{PSINAR(1)} &  Binomial & Power Series & NA & \citet{bourguignon2015first} \\
\hline
\textbf{DP INAR(1)}  & Binomial & double Poisson & NA & \citet{bourguignon2019extended}\\
\hline 
\textbf{GP INAR(1)} &  Binomial & Generalized Poisson & NA &  \citet{bourguignon2019extended} \\
\hline
\textbf{BL-INAR(1)} &  Binomial & Bell & NA & \citet{huang2021new} \\
\hline
\textbf{PL-INAR(1)} &  Binomial & Poisson-Lindley & NA & \citet{livio2018inar} \\
\hline
\textbf{INARKF(1)} &  Binomail & Katz family & NA & \citet{kim2017first} \\
\hline
\textbf{ZP-INAR(1)} &  Binomial & zero-inflated Poisson & NA & \citet{kim2017first} \\
\hline
\textbf{INAR(1)} & Binomial & Lerch, good, weighted Poisson (WP), power-law WP, Poisson polynomial & NA & \citet{weiss2013integer}\\
\hline
\textbf{DCINAR(1)}  &  Generalized Binomial & Mixture of zero and Geometric RVs & Geometric &  \citet{ristic2013geometric}\\
\hline
\textbf{$\rho$-GINAR(1)} & $\rho$-Binomial & Derived to satisfy marginal & Geometric & \citet{borges2016geometric} \\
\hline
\end{tabular}
\end{table}

\begin{table}[p]
\caption{\label{tab:examples2} Examples of GINAR processes found in literature. 'NA' indicates that a closed form expression for the marginal distribution does not exist.}\centering
\hspace*{-0.8cm}
\begin{tabular}{|p{3cm}p{4cm}p{4cm}c p{3cm}|}
\hline
\textbf{Process} & \textbf{Thinning} & \textbf{Innovation} & \textbf{Marginal} & \textbf{Reference} \\
& \textbf{Operator} & \textbf{Dist.} & \textbf{Dist.}& \\
\hline
\textbf{Geometric Minifaction INAR(1)} &  Neg. Binomial & Geometric & Geometric & \citet{aleksic2021geometric} \\
\hline
\textbf{NBTINAR(1)} &  Neg. Binomial & non-negative RV & NA &\citet{wang2021estimation}\\
\hline
\textbf{NonLINAR(1)} &  Geometric & non-negative RV & No closed form &\citet{barreto2023non} \\
\hline
\textbf{Extended Poisson INAR(1)} &  Bernouili-Poisson Convolution & Poisson & NA & \citet{weiss2015poisson} \\
\hline
\textbf{GINARS($p$)} & non-negative RV & Signed Generalized Power Series & NA & \citet{zhang2010inference} \\
\hline
\end{tabular}
\end{table}

\newpage

\section{Proofs of Results}
\label{sec:proofs}
In this section we present proofs for some results presented in the main article.

\textbf{Proof of Lemma 2.1}: For property (c) the result follows from iterated expectations and mean property of the thinning operator where 
\begin{align*}
    E(\alpha \odot X) &= E \left\{E\left(\alpha \odot X|X \right) \right\} \\
    &= E\left\{ \alpha X \right \} \\
    &= \alpha E(X).
\end{align*}

Property (d) also follows from iterated expectation and variance and mean properties of the thinning operator given by
\begin{align*}
    E(\alpha \odot X)^2 &= \var(\alpha \odot X) + (E(\alpha \odot X))^2\\
    &= E(\var(\alpha \odot X)|X) + \var(E(\alpha \odot X|X)) + \alpha^2 (E(X))^2\\
    &= E(X\beta) + \var(\alpha X) + \alpha^2 (E(X))^2 \\
    &= \beta E(X) + \alpha^2 E(X^2) - \alpha^2 (E(X))^2 + \alpha^2 (E(X))^2 \\
    & = \alpha^2 E(X^2) + \beta E(X).
\end{align*}

\textbf{Proof of Lemma 2.2}: The conditional expectation and variance property can be derived as follows: 
\begin{align*}
    E(X_t|X_{t-1},...,X_{t-p}) &= \sum_{j=1}^p E(\alpha_j \odot X_{t-i}|X_{t-1},...,X_{t-p}) + E(\epsilon_t|X_{t-1},...,X_{t-p}) \\
    &= \sum_{j=1}^p E(\alpha_j \odot X_{t-j}|X_{t-j}) + E(\epsilon_t) \\
    &= \sum_{j=1}^p \alpha_j X_{t-j} + \mu_\epsilon,
\end{align*}
and 
\begin{align*}
\var(X_t|X_{t-1},...,X_{t-p}) &= \sum_{j=1}^p \var(\alpha_j \odot X_{t-i}|X_{t-1},...,X_{t-p}) + \var(\epsilon_t|X_{t-1},...,X_{t-p}) \\
&= \sum_{j=1}^p \beta_j X_{t-j} + \sigma^2_\epsilon.
\end{align*}

The marginal mean is obtained through the law of iterated expectations and stationarity of the process as follows:
\begin{align*}
    \mu_X &= E\left( E(X_t|X_{t-1}, \ldots, X_{t-p})\right) \\
    &= E\left(\sum_{j=1}^p \alpha_j X_{t-j} + \mu_\epsilon \right) \\
    &= \sum_{j=1}^p \alpha_j \mu_X + \mu_{\epsilon}.
\end{align*}
Rearranging we get
\begin{align*}
    \mu_X = \frac{\mu_{\epsilon}}{1-\sum_{j=1}^p \alpha_j}.
\end{align*}

The marginal variance of the process is obtained as follows:
\begin{align*}
    \var(X_t) &= \var(E(X_t|X_{t-1},...,X_{t-p})) + E(\var(X_t|X_{t-1},...,X_{t-p})) \\
    &= E\left(\sum_{j=1}^p \beta_j X_{t-j} + \sigma^2_\epsilon \right) + \var\left(\sum_{j=1}^p \alpha_j X_{t-j} + \mu_\epsilon\right) \\
    &= \sum_{j=1}^p \beta_j \mu_X + \sigma^2_\epsilon + \sum_{j=1}^p \sum_{k=1}^{p} \alpha_k \alpha_j   \rho_X(j-k)\var(X_t) \\
    &= \sum_{j=1}^p \beta_j \mu_X + \sigma^2_\epsilon + \sum_{j=1}^p \alpha_j   \rho_X(j)\var(X_t). 
\end{align*}
Note that the last step follows from: 
\begin{align*}
    \var(X_t) \sum_{j=1}^p \alpha_j \alpha_k  \rho_X(j) &= \var\left(\boldsymbol{\alpha}^T \boldsymbol{X}\right)\\
    &= \var(X_t) \boldsymbol{\alpha^T}\boldsymbol{R}\boldsymbol{\alpha} \\
    &= \var(X_t) \boldsymbol{\alpha^T}\boldsymbol{r},
\end{align*}
where $\boldsymbol{R} = [\rho_X(|i-j|)]_{p \times p}, \ \boldsymbol{r} = (\rho_X(1) \dots \rho_X(p))^T, \text{and }  \boldsymbol{\alpha} = (\alpha_1,\dots,\alpha_p)^T.$

Thus
\begin{align*}
    \var(X_t) = \displaystyle\frac{\sum_{j=1}^p \beta_j \mu_X + \sigma^2_\epsilon}{1-\sum_{j=1}^p \alpha_j   \rho_X(j)}.
\end{align*}

Next, we present a proof for the autocorrelation and autocovariance properties of the process. To derive the autocovariance structure we present the GINAR(\textit{p}) model in matrix form similar to \citet{duli1991}. Let $\boldsymbol{X_t} = (X_t, X_{t-1},...,X_{t-p+1})^T$, and let $\boldsymbol{A}$ and $\boldsymbol{\epsilon_t}$ be defined as 
\begin{align*}
    \boldsymbol{A} = \begin{bmatrix}
\alpha_1 & \alpha_2 & \dots & \alpha_{p-1} & \alpha_p \\
1 & 0 & \dots & 0 & 0 \\
\vdots & \vdots &  &\vdots & \vdots \\
0 & 0 & \dots & 1 & 0
\end{bmatrix} \text{and }
\boldsymbol{\epsilon_t} =  \begin{bmatrix}
\epsilon_t \\
0\\
\vdots \\
0
\end{bmatrix}.
\end{align*}
Then the GINAR(\textit{p}) model is given by
\begin{align*}
    \boldsymbol{X_t} = \boldsymbol{A} \odot \boldsymbol{X_{t-1}} + \boldsymbol{\epsilon_t}.
\end{align*}
We define the autocovariance sequence $\{\gamma_X(k): k \in \mathbb{Z}\}$ for $k \neq 0$ by 
\begin{align*}
    \boldsymbol{\gamma_X(k)} = E\big[(\boldsymbol{X_t} - E(\boldsymbol{X_t}))(\boldsymbol{X_{t-k}}-E(\boldsymbol{X_{t-k}}))^T\big].
\end{align*}
Then 
\begin{align*}
    \boldsymbol{\gamma_X(k)} &=  E(\boldsymbol{X_t X_{t-k}^T}) - E(\boldsymbol{X_t})E(\boldsymbol{X_{t-k}})^T \\
    &= E((\boldsymbol{A}\odot\boldsymbol{X_{t-1}+\boldsymbol{\epsilon_t}})\boldsymbol{X_{t-k}}^T) - E(\boldsymbol{X_t})E(\boldsymbol{X_{t-k}})^T \\
    &= \boldsymbol{A}E(\boldsymbol{X_{t-1}}\boldsymbol{X_{t-k}}^T) + E(\boldsymbol{\epsilon_t})E(\boldsymbol{X_{t-k}})^T - E(\boldsymbol{X_t})E(\boldsymbol{X_{t-k}})^T \\
    &= \boldsymbol{A}E(\boldsymbol{X_{t-1}}\boldsymbol{X_{t-k}}^T) - \{E(\boldsymbol{X_t}) - E(\boldsymbol{\epsilon_t})\} E(\boldsymbol{X_{t-k}})^T \\
    &= \boldsymbol{A}\{E(\boldsymbol{X_{t-1}}\boldsymbol{X_{t-k}}^T) - E(\boldsymbol{X_{t-1}})E(\boldsymbol{X_{t-k}})^T\} \\
    &= \boldsymbol{A \gamma(k-1)},
\end{align*}
where we are using the property that $E(\boldsymbol{X_t}) = \boldsymbol{A}E(\boldsymbol{X_{t-1}}) + E(\boldsymbol{\epsilon_t})$ and the shift invariant property of the autocovariance function. Hence it follows that
\begin{align*}
    \gamma_X(k) = \alpha_1 \gamma_X(k-1) + \alpha_2 \gamma_X(k-2)+...+\alpha_p \gamma_X(k-p),
\end{align*}
and consequently the autocorrelation at lag \textit{k} is
\begin{align*}
    \rho_X(k) = \alpha_1 \rho_X(k-1) + \alpha_2 \rho_X(k-2) + ... + \alpha_p \rho_X(k-p).
\end{align*} 
For a proof of the spectral density function property refer to \citet{silva2005difference}. \qed
\newline

\textbf{Proof of Theorem 3.1}:
For the proof we show that the conditions of Theorem 4.1.2 and Theorem 4.1.3 in \citet{amemiya1985advanced} are fulfilled, similar to the proof of \citet{liu2021maximum} for a different INAR process. These theorems give us consistency of the estimator and asymptotic normality. First, we verify the conditions of Theorem 4.1.2 \citep{amemiya1985advanced}. 

(i) By assumption and properties of the GINAR($p$) model, the parameter space \\ $\Theta = \{\boldsymbol{\theta}| \boldsymbol{\theta} = (\alpha_1, \alpha_2,\ldots, \alpha_p, \lambda_{\epsilon}, \sigma^2_{\epsilon}) \}$ is compact. All elements of $\boldsymbol{\theta}$ are bounded by positive constants and $\boldsymbol{\theta_0}$ is an interior point of $\Theta$. Therefore, the parameter space $\Theta$ is an open subset of $\mathbb{R}^k$. 

(ii) Let $P_{\boldsymbol{\theta}}$ be the probability measure under the parameter $\boldsymbol{\theta}$ and assume the following notation:
\begin{align*}
    U_t(\boldsymbol{\theta}) &= \log P_{\boldsymbol{\theta}}(X_t|X_{t-1},X_{t-2}, \ldots, X_{t-p}); \\
    Q_n(\boldsymbol{\theta}) &= \frac{1}{n} \sum_{t=p+1}^n \log P_{\boldsymbol{\theta}}(X_t|X_{t-1},\ldots,X_{t-p})  
    = \frac{1}{n} \sum_{t=p+1}^n U_t(\boldsymbol{\theta}).
\end{align*}
Note that $U_t(\boldsymbol{\theta})$ is continuous for all $\boldsymbol{\theta} \in \Theta$ and hence is a measurable function of the data for all $\boldsymbol{\theta} \in \Theta$. Also note that
\begin{align*}
 \frac{\partial U_t(\boldsymbol{\theta})}{\partial \boldsymbol{\theta}} =  \sum_{t=p+1}^n \frac{\partial P_{\boldsymbol{\theta}}(X_t|X_{t-1},X_{t-2}, \ldots, X_{t-p)}/\partial \boldsymbol{\theta}}{P_{\boldsymbol{\theta}}(X_t|X_{t-1},X_{t-2}, \ldots, X_{t-p})}
\end{align*}
 is clearly continuous in an open neighborhood of $\boldsymbol{\theta}_0$.

(iii) First, note that $U_t(\boldsymbol{\theta})$ is continuous in an open and convex neighborhood of $\boldsymbol{\theta}_0$, denoted by $N(\boldsymbol{\theta}_0)=(\boldsymbol{\theta}_0-\epsilon, \boldsymbol{\theta}_0+\epsilon)$, for all $\epsilon > 0$. This implies the existence of at least one point s.t.
\begin{align*}
    U_t(\Tilde{\boldsymbol{\theta}}) = \underset{\boldsymbol{\theta} \in \Theta}{\arg\max} \
 U_t(\boldsymbol{\theta}).
\end{align*}
Then using Jensen's inequality we can get the following result: 
\begin{align*}
    E\left(\underset{\boldsymbol{\theta \in N(\boldsymbol{\theta}_0)}}{\sup} U_t(\boldsymbol{\theta}) \right) &= E\left( \log U_t(\Tilde{\boldsymbol{\theta}}) \right) \\
    &= E\left( \log P_{\Tilde{\boldsymbol{\theta}}}(X_t|X_{t-1},\ldots,X_{t-p}) \right) \\
    &\leq \log E\left(P_{\Tilde{\boldsymbol{\theta}}}(X_t|X_{t-1},\ldots,X_{t-p}) \right) \\
    &< \infty.
\end{align*}

The conditions for the Uniform Law of Large Numbers (ULLN) \citep{andrews1987ulln} is satisfied and we get
\begin{align*}
    \frac{1}{n} \sum_{t=p+1}^n U_t(\boldsymbol{\theta}) \overset{i.p.}{\rightarrow} E(U_t(\boldsymbol{\theta})) \equiv Q_0(t),
\end{align*}
as $n \rightarrow\infty$.

Next, we want to show that $Q_0(t)$ attains a local maximum at $\boldsymbol{\theta_0}$. We can show that $\boldsymbol{\theta_0}$ uniquely maximizes $E(U_t(\boldsymbol{\theta}))$ by 
\begin{align*}
    E(U_t(\boldsymbol{\theta})) &= E\left( \log P_{\boldsymbol{\theta}}(X_t|X_{t-1},\ldots,X_{t-p}) \right) \\
    &= E\left( \log P_{\boldsymbol{\theta}_0}(X_t|X_{t-1},\ldots,X_{t-p}) \right) - E\left( \log \frac{P_{\boldsymbol{\theta}_0}(X_t|X_{t-1},\ldots,X_{t-p})}{P_{\boldsymbol{\theta}}(X_t|X_{t-1},\ldots,X_{t-p})} \right) \\
    &\leq E\left( \log P_{\boldsymbol{\theta}_0}(X_t|X_{t-1},\ldots,X_{t-p}) \right) = E(U_t(\boldsymbol{\theta}_0)).
\end{align*}

Hence, because we have assumed identifiability and given the argument above, we have that $\boldsymbol{\theta}_0$ uniquely maximizes $E(U_t(\boldsymbol{\theta}))$.

The assumptions of Theorem 4.1.2 of \citet{amemiya1985advanced} are satisfied, implying that the roots of $\displaystyle \frac{\partial Q_n(\boldsymbol{\theta})}{\partial \boldsymbol{\theta}}$ are consistent for $\boldsymbol{\theta}_0$.

Next, we show that the conditions of Theorem 4.1.3 of \citet{amemiya1985advanced} are fulfilled, giving asymptotic normality of the CML estimator, $\boldsymbol{\theta}_{CML}$. We have already shown that assumptions of Theorem 4.1.2 of \citet{amemiya1985advanced} are fulfilled which is also a requirement for Theorem 4.1.3.

(iv) Note that $Q_n(\boldsymbol{\theta})$ is a measurable function of $X_t$ for all $\boldsymbol{\theta} \in \Theta$ and is open and continuous in an open and convex neighborhood, $N(\boldsymbol{\theta_0})$, of $\boldsymbol{\theta_0}$. 

(v) Let 
\begin{align*}
  \ell(\boldsymbol{\theta}) = \sum_{t=p+1}^n U_t(\boldsymbol{\theta}).
\end{align*}
We want to show that
\begin{align*}
    n^{-1}\frac{\partial^2 \ell(\boldsymbol{\theta_{n}^{*}})}{\partial \boldsymbol{\theta}\partial\boldsymbol{\theta}^T}
\end{align*}
converges to a finite non-singular matrix 
\begin{align*}
    \boldsymbol{J}(\boldsymbol{\theta_0}) = \lim_{n \rightarrow \infty} n^{-1}E\left(\frac{\partial^2 \ell(\boldsymbol{\theta_0})}{\partial \boldsymbol{\theta}\partial\boldsymbol{\theta}^T} \right)
\end{align*}
in probability for any sequence $\boldsymbol{\theta^*_n}$ such that $ \boldsymbol{\theta^*_n} \rightarrow \boldsymbol{\theta_0}$ as $n \rightarrow \infty$ and $\boldsymbol{\theta}_{n}^*$. Note that $\boldsymbol{\theta^*_n}$ lies in between $\boldsymbol{\theta_0}$ and $\boldsymbol{\theta}_{CML}$.

In Proposition 2.1 we represent the transition probability using characteristics functions, which are three times differentiable. Hence, we have that all partial derivatives of
\begin{align*}
    \frac{\partial U_t(\boldsymbol{\theta})}{\partial \theta_i}
\end{align*}
exist and are three times continuously differentiable in a neighborhood of $\boldsymbol{\theta}_0$, $N(\boldsymbol{\theta}_0)$. Then, similar to step (iii) we have
\begin{align*}
    E\left(\sup_{\boldsymbol{\theta} \in N(\boldsymbol{\theta_0})} \frac{\partial^2 U_t(\boldsymbol{\theta})}{\partial \theta_i \partial \theta_j} \right) = E\left( \frac{\partial^2 U_t(\Tilde{\boldsymbol{\theta}})}{\partial \theta_i \theta_j} \right) < \infty,
\end{align*}
where $\Tilde{\boldsymbol{\theta}} \in N(\boldsymbol{\theta}_0)$. This is because 
\begin{align*}
    \frac{\partial U_t(\boldsymbol{\theta})}{\partial \boldsymbol{\theta}_i \partial \boldsymbol{\theta}_j}
\end{align*}
exists and is continuous in $N(\boldsymbol{\theta}_0)$ for all $i = 1,2,\ldots$ and $j=1,2,\ldots$. 

The conditions for ULLN is satisfied and we get
\begin{align*}
    \frac{1}{n} \sum_{t=p+1}^n \frac{\partial \ell(\boldsymbol{\theta})}{\partial \boldsymbol{\theta} \partial \boldsymbol{\theta}^T} \overset{i.p.}{\rightarrow} E\left(\frac{\partial \ell(\boldsymbol{\theta})}{\partial \boldsymbol{\theta} \partial \boldsymbol{\theta}^T}\right),
\end{align*}
as $n \rightarrow\infty$ uniformly in $\boldsymbol{\theta} \in N(\boldsymbol{\theta_0})$. And by ULLN consequence  
\begin{align*}
    \frac{1}{n} \sum_{t=p+1}^n \frac{\partial \ell(\boldsymbol{\theta^*_n})}{\partial \boldsymbol{\theta} \partial \boldsymbol{\theta}^T} \overset{i.p.}{\rightarrow} E\left(\frac{\partial \ell(\boldsymbol{\theta^*_n})}{\partial \boldsymbol{\theta} \partial \boldsymbol{\theta}^T}\right),
\end{align*}
when $\boldsymbol{\theta_n^*} \rightarrow \boldsymbol{\theta_0}$ as $n \rightarrow\infty$.

(vi) We have the covariance matrix given by
\begin{align*}
    \cov\left(\frac{\partial U_t(\boldsymbol{\theta_0})}{\partial \boldsymbol{\theta}}  \right) = E\left[\left( \frac{\partial U_t(\boldsymbol{\theta_0})}{\partial \boldsymbol{\theta}} \right) \left( \frac{\partial U_t(\boldsymbol{\theta_0})}{\partial \boldsymbol{\theta}}  \right)^T \right],
\end{align*}
since $E\left(\frac{\partial U_t(\boldsymbol{\theta_0})}{\partial \boldsymbol{\theta}} \right) =0$. Also, because of the ergodicity we have 
\begin{align*}
    n^{-1}\frac{\partial \ell(\boldsymbol{\theta_0})}{\partial \boldsymbol{\theta}} \overset{i.p.}{\rightarrow} E\left(\frac{\partial U_t(\boldsymbol{\theta_0})}{\partial \boldsymbol{\theta}} \right)
\end{align*}
as $n \rightarrow \infty$.
Then, using martingale CLT and Cramer-Wold device we get
\begin{align*}
    n^{-1}\frac{\partial \ell(\boldsymbol{\theta_0})}{\partial \boldsymbol{\theta}} \overset{d}{\rightarrow} N\left(\boldsymbol{0}, \boldsymbol{K}(\boldsymbol{\theta_0}) \right),
\end{align*}
as $n \rightarrow \infty$, where 
\begin{align*}
    \boldsymbol{K}(\boldsymbol{\theta_0}) = \lim_{n \rightarrow \infty} E\left(\frac{\partial \ell(\boldsymbol{\theta_0})}{\partial \boldsymbol{\theta}} \left(\frac{\partial \ell(\boldsymbol{\theta_0})}{\partial \boldsymbol{\theta}}\right)^T \right).
\end{align*}

(vii) Finally we note from \citet{latour1998existence} that the $k^{th}$ order moment of $X_t$ is finite. Thus there exists $W(X_t)$ such that $E(W(X_t)) < \infty$ and 
\begin{align*}
    |\frac{\partial^3 \ln l(\boldsymbol{\theta})}{\partial \theta_i \partial \theta_j \partial \theta_k}| < W(X_t).
\end{align*} 
Then, by Taylor expansion we get
\begin{align*}
    \frac{\partial \ell(\boldsymbol{\theta}_{CML})}{\partial \boldsymbol{\theta}} =     \frac{\partial \ell(\boldsymbol{\theta}_{0})}{\partial \boldsymbol{\theta}} +     \frac{\partial \ell(\boldsymbol{\theta}_{n}^*)}{\partial \boldsymbol{\theta} \partial \boldsymbol{\theta}^T}(\widehat{\boldsymbol{\theta}}_{CML}-\boldsymbol{\theta_0}).
\end{align*}
Then
\begin{align*}
    \sqrt{n}(\widehat{\boldsymbol{\theta}}_{CML} - \boldsymbol{\theta_0}) = -\left[n^{-1}\frac{\partial^2 \ell(\boldsymbol{\theta_n^*})}{\partial \boldsymbol{\theta} \partial \boldsymbol{\theta}^T} \right] \left[n^{-1/2} \frac{\partial \ell(\boldsymbol{\theta_0})}{\partial \boldsymbol{\theta}} \right]. 
\end{align*}
Hence, the conditions of Theorem 4.1.3 of \citet{amemiya1985advanced} are fulfilled. \qed 

\textbf{Proof of Proposition 3.1}:
Note that for $x=1,2,\ldots$
\begin{eqnarray*}
    b(x)
    &=&
\frac{1}{2\pi} \int_{-\pi}^{\pi} Re\!\left(\frac{\phi_{X_t|X_{t-1},\ldots,X_{t-p}}(u) \left[ e^{-iux_t} - e^{-iu(x_t+1)} \right] }{1- e^{-iu}} \right) \,du
   \\ &=&
\frac{1}{2\pi} \int_{-\pi}^{\pi} Re\!\left(\frac{\phi_{X_t|X_{t-1},\ldots,X_{t-p}}(u) e^{-iux_t} ( 1- e^{-iu} )  }{1- e^{-iu}} \right) \,du
   \\ &=&
\frac{1}{2\pi} \int_{-\pi}^{\pi} Re\!\left( \phi_{X_t|X_{t-1},\ldots,X_{t-p}}(u) e^{-iux_t}  \right) \,du
   \\ &=&
\frac{1}{\pi} \int_{0}^{\pi} Re\!\left( \phi_{X_t|X_{t-1},\ldots,X_{t-p}}(u) e^{-iux_t}  \right) \,du.
\end{eqnarray*}
Then
\begin{align*}
    \ell(\boldsymbol{\theta}) = \sum_{t=p+1}^n \log b(x).
\end{align*}
The result of Proposition 3.1 follows directly by taking partial derivatives of $\ell(\boldsymbol{\theta})$.

\section{Additional Simulation Tables}
\label{sec:simulations}
In this section we present additional simulation tables and results. In particular we provide results for estimating the parameters of Po-INAR(1), NB-INAR(1) and Geom-INAR(1) processes when the dependence parameter $\vv{\alpha}$ and innovation mean $\mu_{\epsilon}$ are varied.  We also provide results for Po-INAR(4) and NB-INAR(4) processes when the sample size is varied.

\begin{table}[t]
\caption{\label{tab.inar1.sims.poinar.cl}Estimated bias, SD, and RMSE when estimating the parameters of a Po-INAR(1) process with $\alpha_1=0.5$ and $n=500$. The maximum standard error for each quantity is 0.001 for estimating $\alpha_1$ and 0.031 for estimating $\mu_{\epsilon}$. The number of replicates is 10,000.}
\vspace{.5cm}
\centering
\renewcommand{\arraystretch}{0.7}
\hspace*{-1.3cm}
\begin{tabular}{ll|rr|rr|rr} 
& & \multicolumn{2}{c}{$\mu_{\epsilon}=5$} & \multicolumn{2}{c}{$\mu_{\epsilon}=10$} & \multicolumn{2}{c}{$\mu_{\epsilon}=20$} \\
Method & &  $\widehat{\alpha}_1$ & $\widehat{\mu}_{\epsilon}$ & $\widehat{\alpha}_1$ & $\widehat{\mu}_{\epsilon}$ & $\widehat{\alpha}_1$ & $\widehat{\mu}_{\epsilon}$ \\ 
\hline
{CML}  & Bias & -0.001 & 0.011 & -0.002 & 0.029 & -0.001 & 0.029\\
& SD   & 0.031 & 0.314 & 0.031 & 0.625 & 0.030 & 1.224\\
& RMSE & 0.031 & 0.314 & 0.031 & 0.625 & 0.030 & 1.224\\
\hline {CLS}  & Bias & -0.005 & 0.051 & -0.005 & 0.106 & -0.005 & 0.189\\
& SD   & 0.039 & 0.399 & 0.040 & 0.800 & 0.039 & 1.554\\
& RMSE & 0.039 & 0.399 & 0.040 & 0.800 & 0.039 & 1.554\\
\hline {YW}  & Bias & -0.006 & 0.060 & -0.006 & 0.118 & -0.007 & 0.268\\
& SD   & 0.040 & 0.405 & 0.039 & 0.796 & 0.039 & 1.555\\
& RMSE & 0.040 & 0.405 & 0.039 & 0.796 & 0.039 & 1.555\\
\hline {Pseudo}  & Bias & -0.001 & 0.012 & -0.001 & 0.025 & -0.001 & 0.052\\
& SD   & 0.031 & 0.320 & 0.031 & 0.621 & 0.030 & 1.224\\
& RMSE & 0.031 & 0.320 & 0.031 & 0.621 & 0.030 & 1.224\\
\hline {Whittle}  & Bias & -0.003 & 0.005 & -0.004 & 0.018 & -0.004 & 0.037\\
& SD   & 0.039 & 0.363 & 0.040 & 0.709 & 0.039 & 1.390\\
& RMSE & 0.039 & 0.363 & 0.040 & 0.709 & 0.039 & 1.390\\
\hline {Saddle}  & Bias & -0.011 & 0.111 & -0.007 & 0.129 & -0.004 & 0.161\\
& SD   & 0.031 & 0.313 & 0.030 & 0.613 & 0.031 & 1.226\\
& RMSE & 0.031 & 0.313 & 0.030 & 0.613 & 0.031 & 1.226\\
\end{tabular}
\end{table}

\begin{table}[t]
\caption{\label{tab.inar1.sims.nbinar.mu}Estimated bias, SD, and RMSE when estimating the parameters of a NB-INAR(1) process with $\alpha_1=0.5$, $r=1$ and $n=500$. The maximum standard error for each quantity is 0.001 for estimating $\alpha_1$, 0.004 for estimating $r$  and 0.036 for estimating $\mu_{\epsilon}$. The number of replicates is 10,000.}
\vspace{.5cm}
\centering
\renewcommand{\arraystretch}{0.7}
\begin{tabular}{ll|rrr|rrr|rrr} 
& & \multicolumn{3}{c}{$\mu_{\epsilon}=$ 5} & \multicolumn{3}{c}{$\mu_{\epsilon}=$ 10} & \multicolumn{3}{c}{$\mu_{\epsilon}=$ 20} \\
Method & &  $\widehat{\alpha}_1$  & $\widehat{\mu}_{\epsilon}$  & $\widehat{r}$ & $\widehat{\alpha}_1$ & $\widehat{\mu}_{\epsilon}$ & $\widehat{r}$ & $\widehat{\alpha}_1$ & $\widehat{\mu}_{\epsilon}$ & $\widehat{r}$\\ 
\hline
{CML} \hspace{-1cm} & Bias & -0.002 & 0.020 & -0.002 & -0.001 & 0.009 & 0.000 & 0.000 & 0.025 & 0.000\\
& SD   & 0.021 & 0.314 & 0.136 & 0.017 & 0.574 & 0.118 & 0.014 & 1.062 & 0.105\\
& RMSE & 0.021 & 0.314 & 0.136 & 0.017 & 0.574 & 0.118 & 0.014 & 1.062 & 0.105\\
\hline {CLS}  & Bias & -0.006 & 0.055 & -0.010 & -0.005 & 0.104 & -0.009 & -0.005 & 0.227 & -0.011\\
& SD   & 0.040 & 0.465 & 0.210 & 0.039 & 0.922 & 0.192 & 0.039 & 1.796 & 0.186\\
& RMSE & 0.040 & 0.465 & 0.210 & 0.039 & 0.922 & 0.192 & 0.039 & 1.796 & 0.186\\
\hline {Y-W}  & Bias & -0.007 & 0.067 & -0.014 & -0.007 & 0.130 & -0.013 & -0.006 & 0.242 & -0.009\\
& SD   & 0.040 & 0.475 & 0.205 & 0.039 & 0.914 & 0.192 & 0.039 & 1.797 & 0.187\\
& RMSE & 0.040 & 0.475 & 0.205 & 0.039 & 0.914 & 0.192 & 0.039 & 1.797 & 0.187\\
\hline {Pseudo}  & Bias & -0.005 & 0.049 & -0.009 & -0.006 & 0.115 & -0.010 & -0.005 & 0.200 & -0.008\\
& SD   & 0.040 & 0.462 & 0.196 & 0.039 & 0.919 & 0.188 & 0.039 & 1.815 & 0.184\\
& RMSE & 0.040 & 0.462 & 0.196 & 0.039 & 0.919 & 0.188 & 0.039 & 1.815 & 0.184\\
\hline {Whittle}  & Bias & -0.003 & -0.012 & 0.023 & -0.003 & -0.060 & 0.031 & -0.003 & -0.269 & 0.044\\
& SD   & 0.040 & 0.462 & 0.212 & 0.040 & 0.893 & 0.200 & 0.040 & 1.798 & 0.200\\
& RMSE & 0.040 & 0.462 & 0.212 & 0.040 & 0.893 & 0.200 & 0.040 & 1.798 & 0.200\\
\hline {Saddle}  & Bias & -0.024 & 0.242 & -0.101 & -0.009 & 0.182 & 0.013 & 0.000 & -0.023 & 0.093\\
& SD   & 0.022 & 0.322 & 0.124 & 0.018 & 0.572 & 0.123 & 0.014 & 1.045 & 0.119\\
& RMSE & 0.022 & 0.322 & 0.124 & 0.018 & 0.572 & 0.123 & 0.014 & 1.045 & 0.119\\

\end{tabular}
\end{table}

\begin{table}[t]
\caption{\label{tab.inar1.sims.geom.inar.cl}Estimated bias, SD, and RMSE when estimating the parameters of a Geom-INAR(1) process with $\alpha_1=0.5$ and $n=500$. The maximum standard error for each quantity is 0.001 for estimating $\alpha_1$ and 0.050 for estimating $\mu_{\epsilon}$. The number of replicates is 10,000.}
\vspace{.5cm}
\centering
\renewcommand{\arraystretch}{0.7}
\hspace*{-1.3cm}
\begin{tabular}{ll|rr|rr|rr} 
& & \multicolumn{2}{c}{$\mu_{\epsilon}=5$} & \multicolumn{2}{c}{$\mu_{\epsilon}=10$} & \multicolumn{2}{c}{$\mu_{\epsilon}=20$} \\
Method & &  $\widehat{\alpha}_1$ & $\widehat{\mu}_{\epsilon}$ & $\widehat{\alpha}_1$ & $\widehat{\mu}_{\epsilon}$ & $\widehat{\alpha}_1$ & $\widehat{\mu}_{\epsilon}$ \\ 
\hline
{CML}  & Bias & -0.005 & 0.048 & -0.006 & 0.118 & -0.006 & 0.238\\
& SD   & 0.036 & 0.353 & 0.036 & 0.712 & 0.036 & 1.411\\
& RMSE & 0.036 & 0.353 & 0.036 & 0.712 & 0.036 & 1.411\\
\hline {CLS}  & Bias & -0.006 & 0.057 & -0.005 & 0.099 & -0.006 & 0.222\\
& SD   & 0.042 & 0.415 & 0.040 & 0.798 & 0.040 & 1.588\\
& RMSE & 0.042 & 0.415 & 0.040 & 0.798 & 0.040 & 1.588\\
\hline {YW}  & Bias & -0.006 & 0.061 & -0.007 & 0.129 & -0.006 & 0.229\\
& SD   & 0.041 & 0.409 & 0.040 & 0.793 & 0.040 & 1.587\\
& RMSE & 0.041 & 0.409 & 0.040 & 0.793 & 0.040 & 1.587\\
\hline {Pseudo}  & Bias & -0.006 & 0.053 & -0.006 & 0.109 & -0.006 & 0.236\\
& SD   & 0.038 & 0.379 & 0.037 & 0.729 & 0.036 & 1.415\\
& RMSE & 0.038 & 0.379 & 0.037 & 0.729 & 0.036 & 1.415\\
\hline {Whittle}  & Bias & -0.004 & 0.058 & -0.004 & 0.097 & -0.003 & 0.187\\
& SD   & 0.042 & 0.663 & 0.041 & 1.287 & 0.040 & 2.567\\
& RMSE & 0.042 & 0.663 & 0.041 & 1.287 & 0.040 & 2.567\\
\end{tabular}
\end{table}

\begin{table}[t]
\caption{\label{tab.inar1.sims.poinar.ca}Estimated bias, SD, and RMSE when estimating the parameters of a Po-INAR(1) process with $\mu_{\epsilon}=1$ and $n=500$. The maximum standard error for each quantity is 0.002 for estimating $\alpha_1$ and 0.011 for estimating $\mu_{\epsilon}$. The number of replicates is 10,000.}
\vspace{.5cm}
\centering
\renewcommand{\arraystretch}{0.7}
\hspace*{-1.3cm}
\begin{tabular}{ll|rr|rr|rr} 
& & \multicolumn{2}{c}{$\alpha_1=0.2$} & \multicolumn{2}{c}{$\alpha_1=0.5$} & \multicolumn{2}{c}{$\alpha_1=0.8$} \\
Method & &  $\widehat{\alpha}_1$ & $\widehat{\mu}_{\epsilon}$ & $\widehat{\alpha}_1$ & $\widehat{\mu}_{\epsilon}$ & $\widehat{\alpha}_1$ & $\widehat{\mu}_{\epsilon}$ \\ 
\hline
{CML}  & Bias & -0.002 & 0.002 & -0.001 & 0.002 & -0.001 & 0.004\\
& SD   & 0.045 & 0.069 & 0.032 & 0.073 & 0.014 & 0.074\\
& RMSE & 0.045 & 0.069 & 0.032 & 0.073 & 0.014 & 0.074\\
\hline {CLS}  & Bias & -0.003 & 0.003 & -0.006 & 0.011 & 0.064 & -0.330\\
& SD   & 0.047 & 0.071 & 0.041 & 0.089 & 0.111 & 0.552\\
& RMSE & 0.047 & 0.071 & 0.041 & 0.089 & 0.111 & 0.552\\
\hline {YW}  & Bias & -0.004 & 0.004 & -0.006 & 0.010 & -0.009 & 0.042\\
& SD   & 0.047 & 0.071 & 0.042 & 0.089 & 0.028 & 0.145\\
& RMSE & 0.047 & 0.071 & 0.042 & 0.089 & 0.028 & 0.145\\
\hline {Pseudo}  & Bias & -0.002 & 0.002 & -0.002 & 0.003 & -0.001 & 0.003\\
& SD   & 0.050 & 0.074 & 0.035 & 0.078 & 0.015 & 0.078\\
& RMSE & 0.050 & 0.074 & 0.035 & 0.078 & 0.015 & 0.078\\
\hline {Whittle}  & Bias & -0.001 & 0.000 & -0.003 & 0.002 & -0.005 & 0.007\\
& SD   & 0.047 & 0.085 & 0.042 & 0.082 & 0.029 & 0.081\\
& RMSE & 0.047 & 0.085 & 0.042 & 0.082 & 0.029 & 0.081\\
\hline {Saddlepoint}  & Bias & -0.022 & 0.027 & -0.037 & 0.072 & -0.012 & 0.059\\
& SD   & 0.039 & 0.064 & 0.033 & 0.072 & 0.016 & 0.079\\
& RMSE & 0.039 & 0.064 & 0.033 & 0.072 & 0.016 & 0.079\\
\end{tabular}
\end{table}

\begin{table}[t]
\caption{\label{tab.inar1.sims.nbinar.alpha}Estimated bias, SD, and RMSE when estimating the parameters of a NB-INAR(1) process with $\mu_{\epsilon}=1$, $r=1$ and $n=500$. The maximum standard error for each quantity is 0.005 for estimating $\alpha_1$, 0.012 for estimating $r$  and 2.665 for estimating $\mu_{\epsilon}$. The number of replicates is 10,000.}
\vspace{.5cm}
\centering
\renewcommand{\arraystretch}{0.7}
\begin{tabular}{ll|rrr|rrr|rrr} 
& & \multicolumn{3}{c}{$\alpha_1=$ 0.2} & \multicolumn{3}{c}{$\alpha_1=$ 0.5} & \multicolumn{3}{c}{$\alpha_1=$ 0.8} \\
Method & &  $\widehat{\alpha}_1$  & $\widehat{\mu}_{\epsilon}$  & $\widehat{r}$ & $\widehat{\alpha}_1$ & $\widehat{\mu}_{\epsilon}$ & $\widehat{r}$ & $\widehat{\alpha}_1$ & $\widehat{\mu}_{\epsilon}$ & $\widehat{r}$\\ 
\hline
{CML}  & Bias & -0.002 & 0.002 & -0.003 & -0.002 & 0.003 & -0.007 & -0.001 & 0.004 & -0.008\\
& SD   & 0.036 & 0.075 & 0.192 & 0.029 & 0.080 & 0.214 & 0.014 & 0.083 & 0.238\\
& RMSE & 0.036 & 0.075 & 0.192 & 0.029 & 0.080 & 0.214 & 0.014 & 0.083 & 0.238\\
\hline {CLS}  & Bias & -0.003 & 0.005 & -0.011 & -0.005 & 0.010 & -0.004 & -0.007 & 0.035 & -0.026\\
& SD   & 0.048 & 0.084 & 0.231 & 0.043 & 0.103 & 0.322 & 0.029 & 0.155 & 0.542\\
& RMSE & 0.048 & 0.084 & 0.231 & 0.043 & 0.103 & 0.322 & 0.029 & 0.155 & 0.542\\
\hline {Y-W}  & Bias & -0.004 & 0.004 & -0.008 & -0.007 & 0.014 & -0.018 & -0.009 & 0.044 & -0.033\\
& SD   & 0.048 & 0.085 & 0.231 & 0.042 & 0.102 & 0.308 & 0.029 & 0.155 & 0.541\\
& RMSE & 0.048 & 0.085 & 0.231 & 0.042 & 0.102 & 0.308 & 0.029 & 0.155 & 0.541\\
\hline {Pseudo}  & Bias & -0.005 & 0.006 & -0.018 & -0.006 & 0.011 & -0.017 & -0.005 & 0.025 & -0.032\\
& SD   & 0.059 & 0.094 & 0.230 & 0.043 & 0.102 & 0.275 & 0.024 & 0.130 & 0.410\\
& RMSE & 0.059 & 0.094 & 0.230 & 0.043 & 0.102 & 0.275 & 0.024 & 0.130 & 0.410\\
\hline {Whittle}  & Bias & -0.001 & -0.004 & 0.021 & -0.004 & 0.011 & -0.006 & -0.005 & 0.035 & 0.002\\
& SD   & 0.048 & 0.083 & 0.234 & 0.043 & 0.102 & 0.321 & 0.029 & 0.154 & 0.552\\
& RMSE & 0.048 & 0.083 & 0.234 & 0.043 & 0.102 & 0.321 & 0.029 & 0.154 & 0.552\\
\hline {Saddle}  & Bias & -0.075 & 0.095 & -0.145 & -0.064 & 0.129 & -0.468 & -0.041 & 0.317 & -0.726\\
& SD   & 0.039 & 0.082 & 0.157 & 0.036 & 0.191 & 0.142 & 0.030 & 4.850 & 0.132\\
& RMSE & 0.039 & 0.082 & 0.157 & 0.036 & 0.191 & 0.142 & 0.030 & 4.850 & 0.132\\

\end{tabular}
\end{table}

\begin{table}[t]
\caption{\label{tab.inar1.sims.geominar.ca}Estimated bias, SD, and RMSE when estimating the parameters of a Geom-INAR(1) process with $\mu_{\epsilon}=1$ and $n=500$. The maximum standard error for each quantity is 0.001 for estimating $\alpha_1$ and 0.004 for estimating $\mu_{\epsilon}$. The number of replicates is 10,000.}
\vspace{.5cm}
\centering
\renewcommand{\arraystretch}{0.7}
\hspace*{-1.3cm}
\begin{tabular}{ll|rr|rr|rr} 
& & \multicolumn{2}{c}{$\alpha_1=0.2$} & \multicolumn{2}{c}{$\alpha_1=0.5$} & \multicolumn{2}{c}{$\alpha_1=0.8$} \\
Method & &  $\widehat{\alpha}_1$ & $\widehat{\mu}_{\epsilon}$ & $\widehat{\alpha}_1$ & $\widehat{\mu}_{\epsilon}$ & $\widehat{\alpha}_1$ & $\widehat{\mu}_{\epsilon}$ \\ 
\hline
{CML}  & Bias & -0.005 & 0.005 & -0.006 & 0.008 & -0.007 & 0.017\\
& SD   & 0.045 & 0.068 & 0.042 & 0.078 & 0.031 & 0.105\\
& RMSE & 0.045 & 0.068 & 0.042 & 0.078 & 0.031 & 0.105\\
\hline {CLS}  & Bias & -0.003 & 0.003 & -0.008 & 0.014 & -0.015 & 0.058\\
& SD   & 0.048 & 0.071 & 0.050 & 0.095 & 0.040 & 0.170\\
& RMSE & 0.048 & 0.071 & 0.050 & 0.095 & 0.040 & 0.170\\
\hline {YW}  & Bias & -0.005 & 0.006 & -0.010 & 0.016 & -0.016 & 0.062\\
& SD   & 0.048 & 0.071 & 0.051 & 0.096 & 0.041 & 0.170\\
& RMSE & 0.048 & 0.071 & 0.051 & 0.096 & 0.041 & 0.170\\
\hline {Pseudo}  & Bias & -0.004 & 0.002 & -0.004 & 0.004 & -0.005 & 0.012\\
& SD   & 0.049 & 0.076 & 0.045 & 0.090 & 0.034 & 0.126\\
& RMSE & 0.049 & 0.076 & 0.045 & 0.090 & 0.034 & 0.126\\
\hline {Whittle}  & Bias & -0.002 & -0.001 & -0.007 & 0.017 & -0.011 & 0.062\\
& SD   & 0.049 & 0.106 & 0.050 & 0.152 & 0.041 & 0.228\\
& RMSE & 0.049 & 0.106 & 0.050 & 0.152 & 0.041 & 0.228\\
\end{tabular}
\end{table}

\begin{table}[t]
\caption{\label{tab.inar4.sims.2.poinar.n1}Estimated bias, SD, and RMSE when estimating the parameters of a Po-INAR(4) process with $\alpha_1=0.1$, $\alpha_2 = 0.2$, $\alpha_3 = 0.3$, $\alpha_4 = 0.1$, and $\mu_{\epsilon}=1$. The maximum standard error for each quantity is 0.002 for estimating $\alpha_1$, 0.003 for estimating $\alpha_2$, 0.002 for estimating $\alpha_3$, 0.003 for estimating $\alpha_4$ and 0.014 for estimating $\mu_{\epsilon}$. The number of replicates is 10,000.}
\vspace{.5cm}
\centering
\renewcommand{\arraystretch}{0.7}
\hspace*{-1.3cm}
\begin{tabular}{ll|rrrrr} 
& & \multicolumn{5}{c}{$n=$ 100}\\
Method & &  $\widehat{\alpha}_1$ & $\widehat{\alpha}_2$ & $\widehat{\alpha}_3$ & $\widehat{\alpha}_4$ & $\widehat{\mu}_{\epsilon}$ \\ 
\hline
{CML}  & Bias & 0.001 & -0.017 & 0.003 & -0.024 & 0.118\\
& SD   & 0.089 & 0.104 & 0.089 & 0.104 & 0.493\\
& RMSE & 0.089 & 0.105 & 0.089 & 0.106 & 0.507\\
\hline
{CLS}  & Bias & -0.009 & -0.032 & -0.009 & -0.038 & 0.277\\
&SD   & 0.085 & 0.107 & 0.084 & 0.114 & 0.506\\
&RMSE & 0.085 & 0.111 & 0.084 & 0.120 & 0.576\\
\hline
{Y-W}  & Bias & -0.012 & -0.033 & -0.024 & -0.058 & 0.405\\
& SD   & 0.104 & 0.104 & 0.097 & 0.097 & 0.545\\
& RMSE & 0.105 & 0.109 & 0.100 & 0.114 & 0.679\\
\hline
{Pseudo}  & Bias & -0.002 & -0.025 & -0.003 & -0.031 & 0.175\\
& SD   & 0.091 & 0.107 & 0.088 & 0.106 & 0.448\\
& RMSE & 0.091 & 0.109 & 0.088 & 0.111 & 0.481\\
\hline
{Whittle}  & Bias & 0.010 & -0.016 & 0.003 & -0.043 & 0.099\\
& SD   & 0.090 & 0.102 & 0.088 & 0.102 & 0.459\\
& RMSE & 0.091 & 0.103 & 0.088 & 0.111 & 0.469\\
\end{tabular}
\hspace*{-1cm}
 
\end{table}

\begin{table}[t]
\caption{\label{tab.inar4.sims.2.poinar.n2}Estimated bias, SD, and RMSE when estimating the parameters of a Po-INAR(4) process with $\alpha_1=0.1$, $\alpha_2 = 0.2$, $\alpha_3 = 0.3$, $\alpha_4 = 0.1$, and $\mu_{\epsilon}=1$. The maximum standard error for each quantity is 0.001 for estimating $\alpha_1$, 0.001 for estimating $\alpha_2$, 0.001 for estimating $\alpha_3$, 0.001 for estimating $\alpha_4$ and 0.004 for estimating $\mu_{\epsilon}$. The number of replicates is 10,000.}
\vspace{.5cm}
\centering
\renewcommand{\arraystretch}{0.7}
\hspace*{-1.3cm}
\begin{tabular}{ll|rrrrr} 
& & \multicolumn{5}{c}{$n=$ 500}\\
Method & &  $\widehat{\alpha}_1$ & $\widehat{\alpha}_2$ & $\widehat{\alpha}_3$ & $\widehat{\alpha}_4$ & $\widehat{\mu}_{\epsilon}$ \\ 
\hline
{CML}  & Bias & -0.003 & -0.004 & -0.003 & -0.005 & 0.047\\
& SD   & 0.043 & 0.044 & 0.043 & 0.042 & 0.181\\
& RMSE & 0.043 & 0.044 & 0.043 & 0.043 & 0.187\\
\hline
{CLS}  & Bias & -0.003 & -0.006 & -0.004 & -0.008 & 0.066\\
& SD   & 0.045 & 0.047 & 0.044 & 0.046 & 0.216\\
& RMSE & 0.045 & 0.048 & 0.044 & 0.046 & 0.225\\
\hline
{Y-W}  & Bias & -0.003 & -0.006 & -0.004 & -0.011 & 0.078\\
& SD   & 0.045 & 0.045 & 0.044 & 0.045 & 0.205\\
& RMSE & 0.045 & 0.046 & 0.044 & 0.046 & 0.219\\
\hline
{Pseudo}  & Bias & -0.002 & -0.005 & -0.003 & -0.005 & 0.042\\
& SD   & 0.045 & 0.045 & 0.045 & 0.044 & 0.188\\
& RMSE & 0.045 & 0.046 & 0.045 & 0.044 & 0.192\\
\hline
{Whittle}  & Bias & 0.000 & -0.004 & -0.001 & -0.009 & 0.034\\
& SD   & 0.044 & 0.045 & 0.044 & 0.045 & 0.188\\
& RMSE & 0.044 & 0.045 & 0.044 & 0.046 & 0.191\\
\end{tabular}
\hspace*{-1cm}
 
\end{table}

\begin{table}[t]
\caption{\label{tab.inar4.sims.2.geominar}Estimated bias, SD, and RMSE when estimating the parameters of a Geom-INAR(4) process with $\alpha_1=0.1$, $\alpha_2 = 0.2$, $\alpha_3 = 0.3$, $\alpha_4 = 0.1$, and $\mu_{\epsilon}=1$. The maximum standard error for each quantity is 0.002 for estimating $\alpha_1$, 0.002 for estimating $\alpha_2$, 0.002 for estimating $\alpha_3$, 0.003 for estimating $\alpha_4$ and 0.014 for estimating $\mu_{\epsilon}$. The number of replicates is 10,000. }
\vspace{.5cm}
\centering
\renewcommand{\arraystretch}{0.7}
\hspace*{-1.3cm}
\begin{tabular}{ll|rrrrr} 
& & \multicolumn{5}{c}{$n=$100}\\
Method & &  $\widehat{\alpha}_1$ & $\widehat{\alpha}_2$ & $\widehat{\alpha}_3$ & $\widehat{\alpha}_4$ & $\widehat{\mu}_{\epsilon}$ \\ 
\hline
{CML}  & Bias & -0.008 & -0.032 & -0.010 & -0.049 & 0.310\\
&SD   & 0.082 & 0.096 & 0.078 & 0.099 & 0.482\\
&RMSE & 0.082 & 0.102 & 0.079 & 0.110 & 0.573\\
\hline
{CLS}  & Bias & -0.009 & -0.032 & -0.009 & -0.043 & 0.286\\
& SD   & 0.087 & 0.110 & 0.083 & 0.115 & 0.500\\
& RMSE & 0.087 & 0.115 & 0.084 & 0.123 & 0.576\\
\hline
{Y-W}  & Bias & -0.016 & -0.035 & -0.023 & -0.059 & 0.416\\
& SD   & 0.105 & 0.106 & 0.098 & 0.101 & 0.546\\
& RMSE & 0.106 & 0.112 & 0.101 & 0.117 & 0.686\\
\hline
{Pseudo}  & Bias & -0.008 & -0.030 & -0.010 & -0.044 & 0.261\\
& SD   & 0.085 & 0.102 & 0.083 & 0.106 & 0.522\\
& RMSE & 0.086 & 0.106 & 0.084 & 0.115 & 0.583\\
\hline
{Whittle}  & Bias & 0.011 & -0.018 & 0.003 & -0.044 & 0.128\\
& SD   & 0.092 & 0.104 & 0.087 & 0.105 & 0.584\\
& RMSE & 0.092 & 0.106 & 0.087 & 0.114 & 0.598\\
\end{tabular}
\hspace*{-1cm}
 
\end{table}

\begin{table}[t]
\caption{\label{tab.inar4.sims.3.geominar}Estimated bias, SD, and RMSE when estimating the parameters of a Geom-INAR(4) process with $\alpha_1=0.1$, $\alpha_2 = 0.2$, $\alpha_3 = 0.3$, $\alpha_4 = 0.1$, and $\mu_{\epsilon}=1$. The maximum standard error for each quantity is 0.001 for estimating $\alpha_1$, 0.001 for estimating $\alpha_2$, 0.001 for estimating $\alpha_3$, 0.001 for estimating $\alpha_4$ and 0.004 for estimating $\mu_{\epsilon}$. The number of replicates is 10,000.}
\vspace{.5cm}
\centering
\renewcommand{\arraystretch}{0.7}
\hspace*{-1.3cm}
\begin{tabular}{ll|rrrrr} 
& & \multicolumn{5}{c}{$n=$ 500}\\
Method & &  $\widehat{\alpha}_1$ & $\widehat{\alpha}_2$ & $\widehat{\alpha}_3$ & $\widehat{\alpha}_4$ & $\widehat{\mu}_{\epsilon}$ \\ 
\hline
{CML}  & Bias & -0.003 & -0.007 & -0.002 & -0.008 & 0.066\\
& SD   & 0.043 & 0.044 & 0.042 & 0.044 & 0.189\\
& RMSE & 0.043 & 0.045 & 0.042 & 0.045 & 0.200\\
\hline
{CLS}  & Bias & -0.003 & -0.007 & -0.003 & -0.009 & 0.067\\
& SD   & 0.045 & 0.048 & 0.045 & 0.048 & 0.213\\
& RMSE & 0.045 & 0.049 & 0.045 & 0.049 & 0.223\\
\hline
{Y-W}  & Bias & -0.001 & -0.007 & -0.005 & -0.014 & 0.081\\
& SD   & 0.046 & 0.047 & 0.045 & 0.047 & 0.207\\
& RMSE & 0.046 & 0.047 & 0.045 & 0.049 & 0.222\\
\hline
{Pseudo}  & Bias & -0.002 & -0.006 & -0.004 & -0.008 & 0.054\\
& SD   & 0.046 & 0.048 & 0.046 & 0.047 & 0.209\\
& RMSE & 0.046 & 0.048 & 0.046 & 0.048 & 0.216\\
\hline
{Whittle}  & Bias & 0.001 & -0.005 & -0.001 & -0.010 & 0.039\\
& SD   & 0.045 & 0.048 & 0.045 & 0.048 & 0.236\\
& RMSE & 0.045 & 0.048 & 0.045 & 0.049 & 0.239\\

\end{tabular}
\hspace*{-1cm}
 
\end{table}

\clearpage
\small
{\renewcommand{\baselinestretch}{1.1}\setlength{\bibsep}{2pt}
\small\bibliographystyle{abbrvnat}\bibliography{supplement}}